\documentclass[prl,amsmath,amssymb,superscirptaddress,reprint,floatfix]{revtex4-1}

\usepackage{graphicx}
\graphicspath{{./SagnacGraphics/SagnacFigures/}}
\usepackage{bm}
\usepackage{amsmath,amssymb}
\usepackage{color}
\usepackage[caption=false]{subfig}

\newcommand{\db}{d_{\mathrm{B}}}
\newcommand{\rb}{r_{\mathrm{B}}}

\newcommand{\real}{\mathrm{Re}}
\newcommand{\imag}{\mathrm{Im}}

\newcommand{\chib}{\chi_1}
\newcommand{\chieit}{\chi_0}

\newcommand{\rc}{\rho}

\newcommand{\FF}{\mathcal{F}}

\newcommand{\BSS} {\begin{pmatrix} 1 & i \\ i & 1 \end{pmatrix}}
\newcommand{\Evec}{\vec{E}}
\newcommand{\M}{ \begin{pmatrix} \sigma & \kappa \\  -\kappa &  -\sigma \end{pmatrix} }
\newcommand{\F}{\frac{1}{\lambda} \begin{pmatrix} \lambda\cos(\lambda z) - i\sigma\sin(\lambda z) & -i\kappa\sin( \lambda z) \\ i\kappa\sin(\lambda z) & \lambda\cos (\lambda z) + i\sigma\sin(\lambda z)  \end{pmatrix} }
\newcommand{\rss}{u}

\newcommand{\ks}{k_\mathrm{s}}
\newcommand{\deltat}{\delta_{\mathrm{mod}}}
\begin{document}

\title{Induced cavities for photonic quantum gates}

\author{Ohr Lahad}
\affiliation{Department of Physics of Complex Systems, Weizmann Institute of Science, Rehovot 76100, Israel}
\author{Ofer Firstenberg}
\affiliation{Department of Physics of Complex Systems, Weizmann Institute of Science, Rehovot 76100, Israel}

\date{\today}

\begin{abstract}
Effective cavities can be optically-induced in atomic media and employed to strengthen optical nonlinearities. Here we study the integration of induced cavities with a photonic quantum gate based on Rydberg blockade. Accounting for loss in the atomic medium, we calculate the corresponding finesse and gate infidelity. Our analysis shows that the conventional limits imposed by the \emph{blockade} optical depth are mitigated by the induced cavity in long media, thus establishing the \emph{total} optical depth of the medium as a complementary resource.
\end{abstract}

\maketitle

Optical nonlinearities at the few-photon level, manifested by effective strong interactions between individual photons, provide a platform for investigating correlated photonic states \cite{Carusotto2013,Otterbach2013,Shahmoon2014,Gorshkov2015,Firstenberg2013AttrPhot} and enable optical quantum computing and networks \cite{OBrien2007,OBrien2009,Kimble2008}. The effective interaction between photons is mediated by strongly coupling them to single quantum emitters or to ensembles of cooperating emitters \cite{Chang2014,FirstenbergRMP2017}.
When employing single atoms, strong coupling is obtained using high-finesse optical cavities \cite{Vuletic2011,Kimble2004,Rempe2016Gate,Lukin2014PC,Rauschen2014Pi}. Cooperating ensembles, namely interacting Rydberg atoms, can reach the strong coupling regime without a cavity \cite{FirstenbergReviewJPB2016,Peyronel2012,Durr2016,Tresp2016Absorber,AdamsReview2013}.

\begin{figure}[b] 
\includegraphics[scale=0.31,trim=0.cm 20.0cm 0.cm 0.0cm,clip=true]{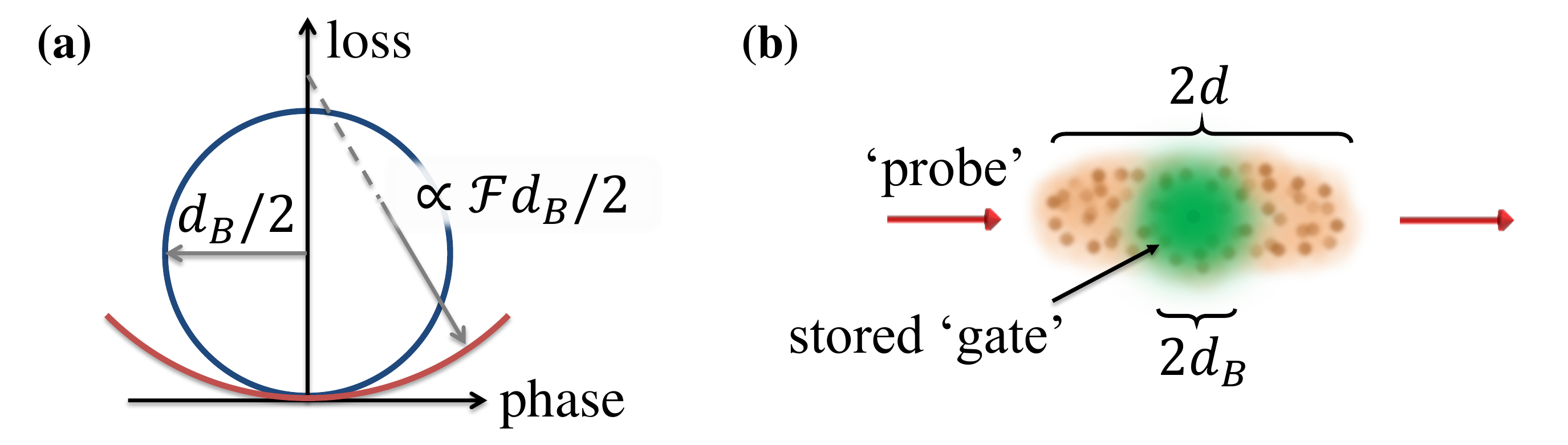}
\caption{(a) When light traverses an atomic medium and acquires a nonzero phase, it always experiences some loss. Given the resonant optical depth $2\db$, the blue circle traces the relation between phase and loss for two-level atoms or under the conditions of ideal EIT [Eq.~(\ref{eq_circle}), see text].
In a cavity, the radius of the circle effectively grows with the cavity finesse $\FF$ (red). (b) A phase gate based on Rydberg blockade by a stored photon.
}
\label{fig1p1}
\end{figure}

The cooperativity of Rydberg atoms stems from a blockade mechanism due to strong Rydberg-Rydberg interactions \cite{Lukin2001,Saffman2010RydbergRev,PritchardPRL2010}. Within the so-called blockade volume, the narrow-band optical excitation of Rydberg states is limited to one collective state.
Consequently, the blockade volume functions as a ``superatom" with a cross-section enhanced by the large number of blockaded atoms \cite{vuletic2006superatom,Saffman2008,Dudin2012}.
The optical depth $2\db$ of the blockade volume is the key parameter determining the strength of the optical nonlinearity. For quantum nonlinear optics, high-fidelity operation of photonic gates requires $\db\gg 1$ \cite{PeyronelNature2012,Firstenberg2013AttrPhot,Gorshkov2013,Adams2014,Moos2015,murray2016many,Durr2016,Tresp2016Absorber,Pohl2017},
with the fundamental limit imposed by the relation between dissipation and dispersion near resonance, see Fig.~\ref{fig1p1}(a).
Unfortunately, the present record $2\db=12.5$ \cite{Tresp2016Absorber} limits the fidelity to $\sim 50\%$ and is difficult to surpass
\cite{Pfau2014Density,DurrPRL2014}.

It has been debated whether the limit imposed by $\db$ can be circumvented in long media, utilizing their large total optical depth $2d\gg 2\db$, see Fig.~\ref{fig1p1}(b). 
For example, two simultaneous photons co-propagating along several blockade volumes may have longer effective interaction time, but the overall fidelity is undermined by spatial entanglement between the pulses and by the narrow transmission bandwidth in long media \cite{murray2016many,Gorshkov2013,Hanspeter2016Kerr,Firstenberg2013AttrPhot,Peyronel2012,LesanovskyPRA2015}. This Letter provides a positive answer to this longstanding question. We show that the Rydberg-mediated interaction can be strengthened by utilizing long media as effective cavities, whose finesse $\FF$ grows as the square root of the total optical depth $2d$.

\begin{figure}[bt] 
\includegraphics[width=8.6cm,trim=0.cm 0.8cm 0.cm 1.15cm,clip=true ]{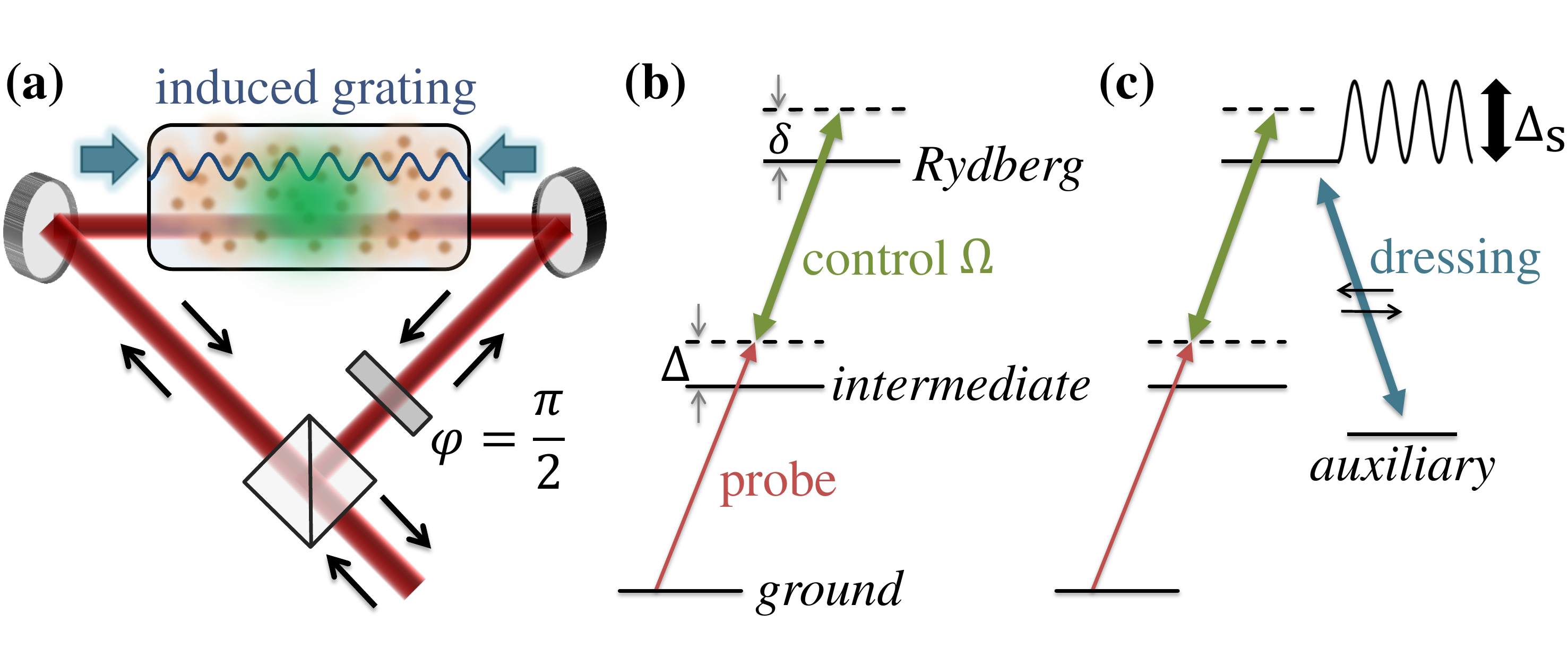}
\caption{(a) Optically-induced grating in a Sagnac interferometer. (b) Atomic level scheme.
A control field (green) couples the probe transition (red) to a Rydberg state, rendering EIT.
(c) To induce a cavity, the EIT resonance frequency is longitudinally modulated by a far-detuned dressing standing-wave (blue).
For example, this scheme can be implemented with rubidium atoms using a probe, control, and dressing fields at 780, 479, and 475 nm, respectively. The angles between the optical axis and the dressing beams are tuned to form a standing wave with a period that is half the probe wavelength.
}\label{fig3}
\end{figure}

We employ a standing-wave dressing field to imprint a Bragg grating in the medium and induce an optical bandgap. This scheme was originally proposed for enhancing nonlinear effects via dynamical control of the bandgap \cite{Andre2002e,Andre2005e}. We follow Hafezi \emph{et al.}~\cite{Hafezi2012} and exploit the transmission resonance outside the bandgap, where the Rydberg-mediated interaction is enhanced without dynamical control.
The enhancement we find is similar to that obtained with \emph{actual} cavities \cite{Grangier2012,SimonPRA2015,Grangier1997,Vuletic2011,GorshkovPRA2007}: the blockaded optical depth effectively experienced by the circulating photons is given by $\FF \db$ \cite{Sorenso2016RydCav,Gong2015}, see Fig.~\ref{fig1p1}(a). 
To render a system with single input and output ports, as required for high-fidelity gate operation, we introduce a Sagnac configuration [Fig.~\ref{fig3}(a)].
While the maximal finesse of optically-induced cavities scales $\propto d$ \cite{Hafezi2012}, we show that the
overall performance of a quantum phase-gate improves approximately $\propto \sqrt{d}$ when accounting for dissipation. Very recently, similar results were reported for so-called ``stationary light" in the strong-coupling (non-cooperative) regime \cite{SorensonArxiv2016Sagnac}.

\emph{Rydberg phase gate.}---We analyze a gate model based on photon storage \cite{Gorshkov2011a,Sorenso2016RydCav,Gong2015,LesanovskyPRA2015,Durr2016}, illustrating the limitations posed by small $\db$ and the resolution offered by a cavity. Here a propagating `probe' photon acquires the phase $\phi=\pi$ conditional on the storage of a `gate' photon in the medium [Fig.~\ref{fig1p1}(b)].
Electromagnetically-induced transparency (EIT) \cite{Fleischhauer2005EITRev} in the ladder arrangement shown in Fig.~\ref{fig3}(b) is employed. First, EIT is used to store the gate photon as a collective excitation comprising one Rydberg atom. 
Afterwards, the probe photon traverses the medium via EIT utilizing a different Rydberg level.

We use the subscripts $j=0,1$ to denote the cases without ($j=0$) and with ($j=1$) the stored Rydberg excitation. In the first case, the probe photon experiences the EIT susceptibility $\chieit$. In the second case, within a blockade radius $\rb$ around the stored excitation, the EIT conditions are violated due to the Rydberg-Rydberg interaction \cite{Raithel2007}, and the probe photon experiences the bare susceptibility of a two-level atom $\chib$. To describe the dynamics of a single probe photon, it is sufficient to consider the linear suceptibilities of the medium. A conditional phase gate is thus obtained when $\phi=\real[\chib-\chieit]k\rb=\pi$, with $k$ the optical wave-vector.

To simplify the discussion, we include no decay of the Rydberg excitation, assuming it is negligible compared to the power broadening $|\Omega^2/(\Delta+i\Gamma)|$, with $2\Omega$ the Rabi frequency of the classical control field, $2\Gamma$ the decay rate of the intermediate state, and  $\Delta$ the detuning from the intermediate state. The susceptibilities then acquire the form \cite{Paspalakis2002} $k\rb\chi_j=-\db\Gamma/\left[i\Gamma + \Delta -\left(1- j\right)\Omega^2/\delta\right]$ for $j=0,1$, where $\db$  is the optical depth over the blockade radius $\rb$, and $\delta$ is the two-photon detuning from the Rydberg state. We observe that $k\rb\chi_j$ satisfy the relation
\footnote{\label{note1} In fact, all multi-level EIT susceptibilities $\chi=\chib [1-\sum_n \Omega_n^2/ (\delta_n+i\gamma_n)/(\Delta+i\Gamma)]^{-1}$ with multiple Rabi frequencies $\Omega_n$ and detunings $\delta_n$ \cite{Paspalakis2002} form the same circle when $\gamma_n=0$.}
\begin{equation}\label{eq_circle}
|k\rb\chi_j-i \db/2|=\db/2~~~~\textrm{for both}~j=0,1,
\end{equation}
forming identical circles in the complex plane, see Fig.~\ref{fig1p1}(a).
Therefore loss of the probe photon ($\propto \imag\chi_j)$ is unavoidable whenever $\phi\ne 0$, which limits the gate fidelity.
The operating point that minimizes the loss has an elegant solution when the whole medium is blocked ($d=\db$ \cite{Tresp2016Absorber}). Then, the loss is quantified by the mean absorption with and without the stored Rydberg excitation $\epsilon=\imag[\chib+\chieit]k\rb$, and we observe that $\phi$ and $\epsilon$ form a circle too, now with twice the radius $|\phi+i\epsilon-i\db|=\db$. It follows that $\phi=\pi$ requires $\db\ge\pi$; For $\db\gg \pi$, the loss $\epsilon=\pi^2/(2\db)$ scales inversely with $\db$.

\begin{figure}[t] 
\includegraphics[width=8.6 cm]{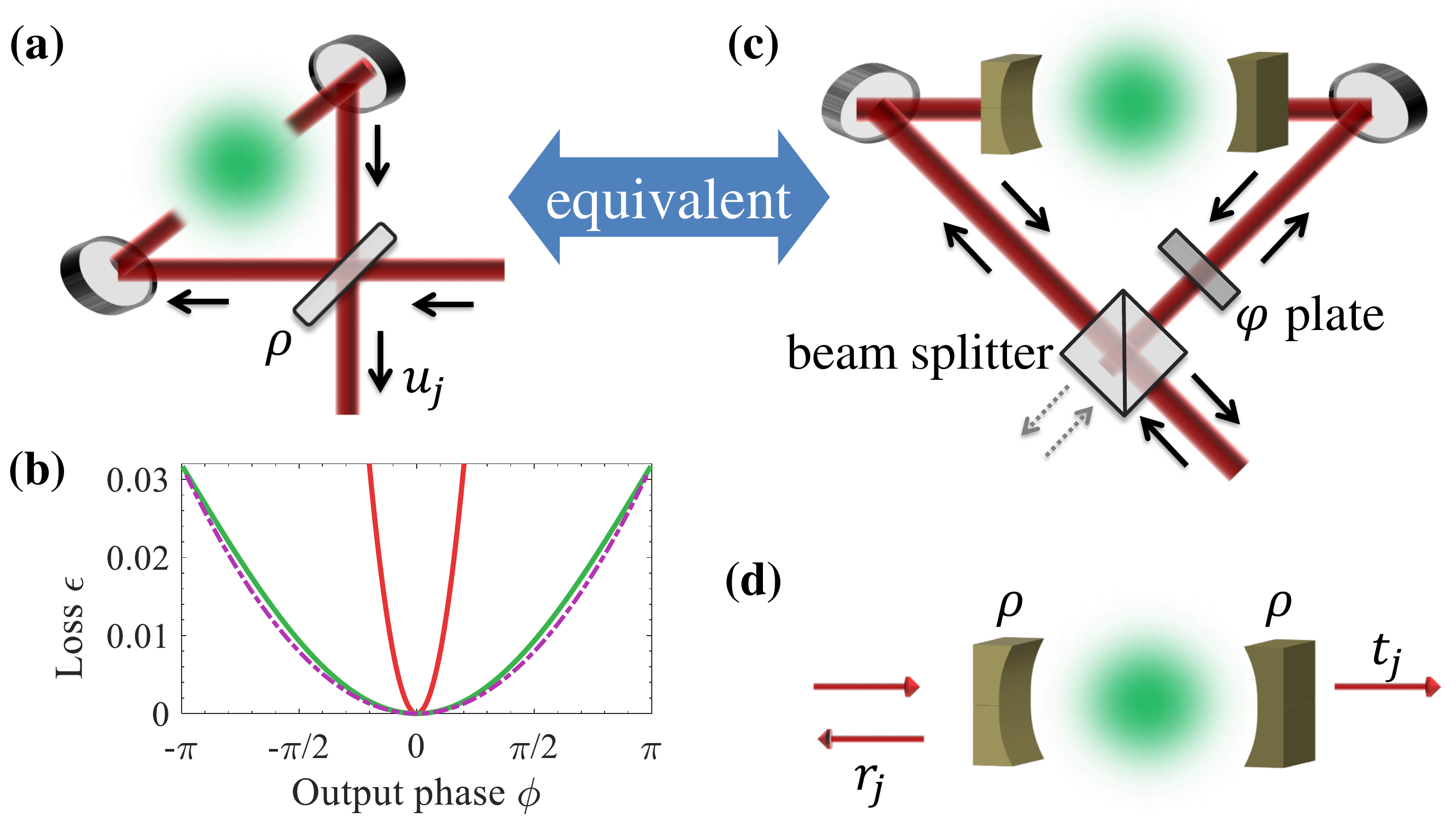}
\caption{Cavities encircling atomic media. (a) Ring cavity.
(b) The loss versus phase, defined in  Eq.~(\ref{eqn:phi_eps}), for the ring cavity ($\FF=20\pi$, $2\db=4\pi$). The exact relation in Eq.~(\ref{eqn:ring_error}) (green) is well approximated by a circle (dashed purple) with a radius $\pi\FF/8$ larger than that obtained with no cavity (red).
(c) A symmetric Fabry-P\'{e}rot cavity inside a Sagnac interferometer is equivalent to a ring cavity. (d) Fabry-P\'{e}rot cavity.
}\label{fig1p2}
\end{figure}

The above limitation can be overcome by incorporating an optical cavity with single input and output ports, such as single-side cavities \cite{Kimble2004,Waks2006,Lukin2014PC, Rempe2016Gate,Sorenso2016RydCav,Gong2015} or ring cavities \cite{Rauschen2014Pi,Dayan2014,Pfau2016Ring}.
For example, consider a ring cavity of length $l$ containing the atomic medium [Fig.~\ref{fig1p2}(a)]. With $\rc$ the reflection amplitude of the coupling mirror, the cavity output amplitude is given by $\rss_j=(\rc +e^{i\theta_j})/(1+\rc e^{i\theta_j})$ \cite{Yariv2007}. The (complex) phase acquired along the ring $\theta_j=kl+k\rb\chi_j$ includes the medium response with ($j=1$) and without ($j=0$) the stored Rydberg excitation. For a bare cavity tuned to resonance $kl=\pi$, gating the medium between $\chieit$ and $\chib$ shifts the cavity across its resonance and alters the reflected phase from $-\pi/2$ to $\pi/2$. Minimal absorption from the medium is obtained at the bottom of the circles described by Eq.~(\ref{eq_circle}), where $\imag[k\rb\chi_j]\approx\real[k\rb\chi_j]^2/\db$. Substituting this into $\theta_j$ and $\rss_j$, and defining again the loss and conditional phase
\begin{equation}\label{eqn:phi_eps}
\epsilon=-\log (|\rss_1|) - \log (|\rss_0|), ~~ \phi=\arg(\rss_1)-\arg(\rss_0),
\end{equation}
we find the relation
\begin{equation}\label{eqn:ring_error}
\epsilon=(4\pi)/(\FF\db)[1-\cos(\phi/2)],
\end{equation}
where the finesse $\FF=\pi(1+\rho)/ (1-\rho )$. For $\phi=\pi$ and comparing to $\epsilon=\pi^2/(2\db)$ obtained without a cavity, we find that $\db$ is effectively increased by the factor $\pi\FF/8$, see Fig.~\ref{fig1p2}(b).

A cavity induced by uniformly dressing the medium, as we shall analyze, is akin to a symmetric Fabry-P\'{e}rot cavity, with \emph{two} pairs of input-output ports. To recover a single-port configuration, we place the two-port cavity inside a Sagnac interferometer, as depicted in Fig.~\ref{fig1p2}(c). The incoming light impinges on the cavity from both sides, with the relative phase tuned by a phase plate $\varphi$.

The transmission matrix from the external input ports to the external output ports of the Sagnac beam splitter (BS) is calculated from
\begin{equation} \label{eqn:SagnacMat}
\frac{1}{2}
\overbrace{  \BSS  } ^ {\mathrm{ BS \, out }  }
\overbrace { \begin{pmatrix}  1 & 0 \\ 0 & e^{i\varphi} \end{pmatrix} } ^ {\varphi\mathrm{-plate} }
\overbrace{ \begin{pmatrix} 0 & 1 \\ 1 & 0 \end{pmatrix} }  ^{\mathrm { flip \, modes} }
\overbrace {   \begin{pmatrix} t_j & r_j \\ r_j & t_j \end{pmatrix}      }   ^ { \mathrm {cavity} }
  \overbrace { \begin{pmatrix}  1 & 0 \\ 0 & e^{i\varphi} \end{pmatrix} } ^ {\varphi\mathrm{-plate} }
 \overbrace{  \BSS }^{\mathrm{BS \, in}},
\end{equation}
with the transmission and reflection amplitudes of the bare cavity [see Fig.~\ref{fig1p2}(d)] given by \cite{Siegman1986}
\begin{equation} \label{eqn:cavityAmps}
t_j=\frac{(1-\rc^2)e^{i\theta_j}}{1+\rc^2e^{2i\theta_j}} \textrm{~~and~~} r_j=i\frac{\rc \left( 1+e^{2i\theta_j} \right) }{1+\rc^2e^{2i\theta_j}}.
\end{equation}
By choosing $\varphi=\pi/2$, the matrix (\ref{eqn:SagnacMat}) becomes diagonal, and the light is back reflected to the port it arrived from. The output amplitude for the first port is $r_j-t_j$, which exactly equals $i\rss_j$ of the ring cavity (with the shift $\theta_j \mapsto \theta_j - \pi/2$). Therefore, while the phase shift of a bare two-port cavity is limited to $\pi/2$ per port around the resonance \footnote{For example, $|\arg(r_0)-\arg(r_1)|< \pi/2$ and $|\arg(t_0)-\arg(t_1)|< \pi/2$ within the resonance linewidth $|\real(\theta_0-\theta_1)|<\pi/\FF$.},
the Sagnac setup enables a $\pi$ conditional phase shift, regaining the single-port properties. We stress that the Sagnac interferometer does not form another cavity, and field build-up occurs only inside the cavity.

\begin{figure} 
\includegraphics[scale=0.47,trim=0cm 0cm 0cm 0cm,clip=true]{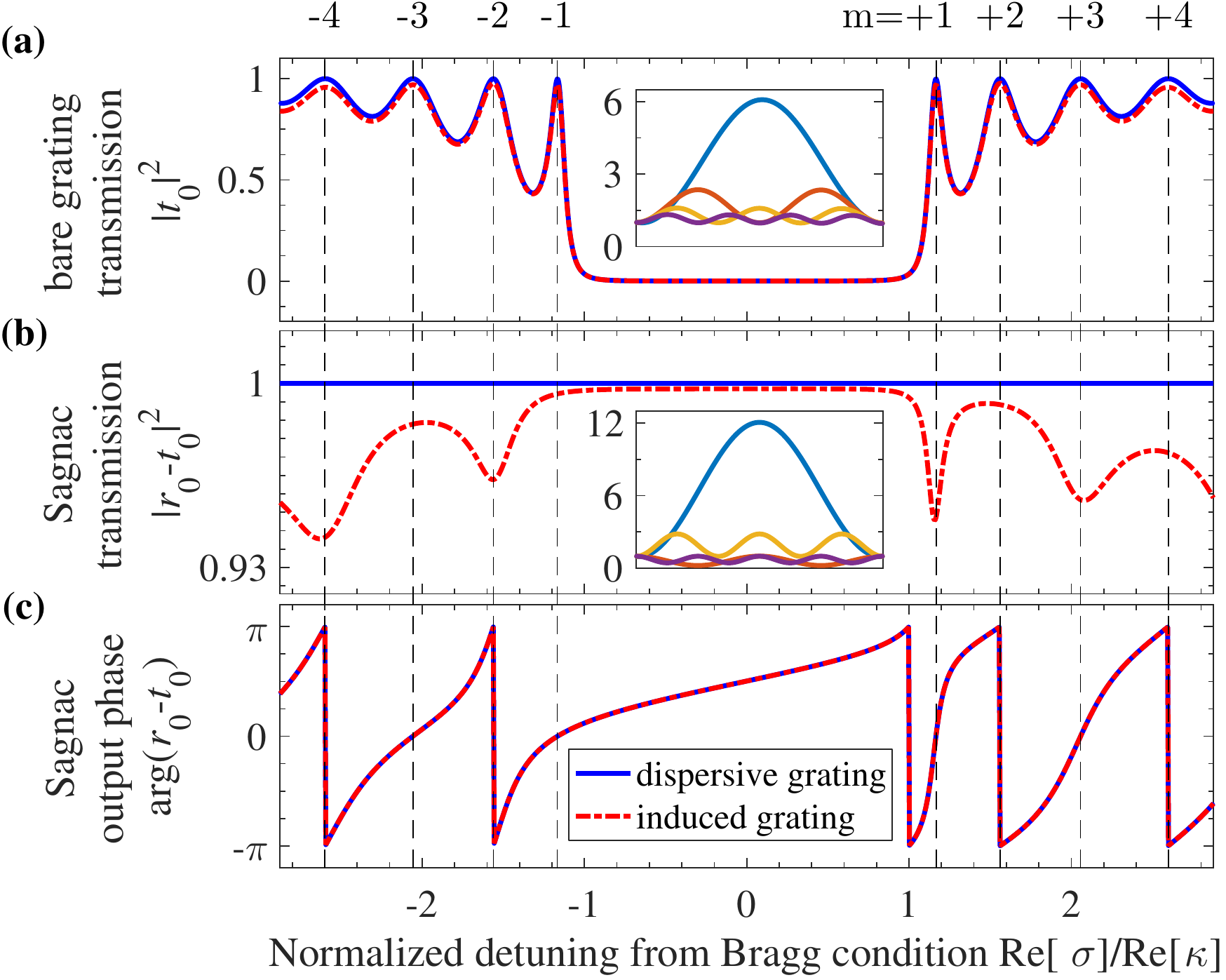}
\caption{Transmission spectra of (a) bare Bragg grating, and (b,c) grating inside a Sagnac interferometer. We compare (solid blue) a purely dispersive Bragg grating ($\kappa L=5.24$, $\imag~\sigma=0$) with (dashed red) a grating formed by dressing an atomic medium ($d=10^4$, and the dressing parameters are chosen to minimize the gate loss when $2\db=4\pi$). The resonances of the bare grating are $m=\pm\left(1,2,3,4...\right)$, but only $m=1,-2,3,-4,...$ retain the steep phase slope in the Sagnac setup. Insets: longitudinal intensity profiles at the corresponding resonances $\pm m=1-4$ (blue, red, orange, purple).
}\label{fig2}
\end{figure}

\emph{Bragg grating.}---We now turn to consider cavities formed by finite media with a uniform longitudinal modulation. A modulation on the wavelength-scale couples the right ($+$) and left ($-$) propagating modes $E_\pm(z)$ according to
\begin{equation}   \label{eqn:gratingProp}
\frac{\partial \Evec(z)}{\partial z} = i \M \Evec(z),~~~\textrm{where}~~\Evec(z)={\begin{pmatrix} E_+(z)  \\  E_-(z) \end{pmatrix}}.
\end{equation}
For example, in a Bragg grating with a modulated susceptibility $\chi(z) =\chi_{\mathrm{DC}}+\chi_{\mathrm{AC}} \cos \left( 2\ks z \right) $, the coupling matrix elements for a probe with wave-number $k$ are given by $\sigma=k-\ks + k\chi_\mathrm{DC}/2$ and $\kappa=k \chi_\mathrm{AC}/4$ \cite{Erdogan1997}.
The solution of Eq.~(\ref{eqn:gratingProp}) can be written for a uniform grating of length $L$ as $\Evec(z)=\textbf{F}_{L-z}\Evec(L)$, where
\begin{equation}
\textbf{F}_z=\F, \nonumber
\end{equation}
with the eigenvalues $\pm \lambda=\pm \sqrt{\sigma^2-\kappa^2}$.
For a field incoming at $z=0$, we substitute $E_+(L)=1$ and $E_-(L)=0$ and obtain the transmission and reflection coefficients, $t_0=1/E_+(0)$ and $r_0=E_-(0)/E_+(0)$. The transmission spectrum, shown in Fig.~\ref{fig2}(a) for a specific set of parameters, exhibits a wide reflection `bandgap' at $\sigma\approx 0$ and narrow transmission resonances around it. These resonances arise due to the finite length of the medium and correspond to the oscillations of $\textbf{F}_z$ outside the bandgap, where the eigenvalues $\pm \lambda$ are predominantly real.
The intensity profile along the medium $\| \Evec (z)\|^2 = |E_+^2|+|E_-^2|$ at the first four resonances $\real(\lambda L)\approx m\pi~(m=1-4)$ is shown in the inset of Fig.~\ref{fig2}(a).
The intensity builds up in the bulk, similarly to a cavity resonance
\cite{Erdogan1997, Hafezi2012}.

The limitation discussed earlier for Fabry-P\'{e}rot cavities applies here as well -- the symmetric two-port cavity formed by the uniform grating cannot alone perform an efficient conditional $\pi$-phase operation. We thus invoke the Sagnac setup, as depicted in Fig.~\ref{fig3}(a). Then the overall output amplitude calculated from (\ref{eqn:SagnacMat}) is $r_0-t_0$, and the field in the bulk is
\begin{equation}
\Evec_\mathrm{Sag}(z)=\frac{1}{\sqrt{2}}\left[ \Evec(z) - \begin{pmatrix} 0 & 1 \\ 1 & 0 \end{pmatrix} \Evec(L-z) \right].
\end{equation}
Fig.~\ref{fig2}(b,c) shows the resulting spectrum and the intensity profiles at the first four resonances. Evidently, only the resonances $m=+1,-2,+3,-4,...$, having anti-symmetric modes, are enhanced in the Sagnac setup. At these resonances, the slope of the output phase and the intensity build-up $B=\max_z\|\Evec_\mathrm{Sag}(z)\|^2>1$ scale linearly with the effective finesse $\FF \approx \pi B$.
The strongest resonance is obtained for $m=+1$, where $\lambda L=\pi(1+i \alpha)$.
Up to first order in the loss $\alpha \ll 1$, we find the overall transmissivity $T$ and finesse $\FF$
\begin{equation} \label{eqn:sagTnB}
T=|r_0-t_0|^2\approx \left| \frac{1- \alpha \kappa L }{1+\alpha\kappa L q} \right|^{2}
\textrm{~~and~~}
\FF  \approx   \frac{1}{\pi}  \left|\frac{(1+q)\kappa L}{1+\alpha\kappa L q} \right|^2,
\end{equation}
where $q=\sqrt{1+\pi^2/(\kappa L)^2}.$

\emph{Induced cavities in atomic media.}---The atomic scheme we analyze in order to optically induce a grating is shown in Fig.~\ref{fig3}(c). A standard Rydberg EIT system as outlined earlier is coupled to an auxiliary atomic state by a dressing standing-wave with wavevector $\ks\approx k$ \cite{Andre2002e}.
The far detuned dressing field gives rise to a longitudinally-periodic light shift of the Rydberg state, effectively modulating the two-photon detuning $\deltat(z)=\delta+\Delta_s \cos \left(2 \ks z \right)$. Staying well within the EIT linewidth $|\deltat|\ll|\Omega^2/(\Delta+i\Gamma)|$, the EIT susceptibility $kL\chieit=-2d\Gamma/(i\Gamma + \Delta - \Omega^2/\deltat)$ can be expanded as,
\begin{equation}   \label{eqn:NSindexMod}
\chieit(z) \approx \frac {2d} {kL} \frac {\deltat (z)}{\Omega^2/\Gamma} \left[ 1  + i   \frac {\deltat (z)}{\Omega^2/(\Gamma-i\Delta)}  \right].
\end{equation}
We now substitute $2\cos^2 (2 \ks z)=1+\cos (4 \ks z)$ and neglect the fast oscillating term $\cos (4 \ks z)$ \cite{Zimmer2006,Andre2002e}.
The terms proportional to $\cos (2 \ks z)$ are identified as $\chi_\mathrm{AC}$, and the rest comprise $\chi_\mathrm{DC}$. Finally, the Bragg coupling coefficients are given by \cite{Erdogan1997}
\begin{flalign}
\sigma=&\Delta k + k \frac{\chi_\mathrm{DC}}{2}=\Delta k  +\frac{d}{L} \left[  x+ (x^2+ 2 y^2)\left(i+\frac{\Delta}{\Gamma}\right) \right]     \nonumber \\
\kappa=&k \frac{\chi_\mathrm{AC}}{4}=\frac{d}{L} \left[  y+ 2x y\left(i+\frac{\Delta}{\Gamma}\right) \right],
\end{flalign}
where $x=\delta\Gamma/\Omega^2$, $y=\Delta_s\Gamma/(2\Omega^2)$, and $\Delta k=k-k_s$. The imaginary parts of $\sigma$ and $\kappa$ account for loss, absent in an ideal Bragg grating.

\begin{figure}[tb]
\includegraphics[scale=0.32,trim=0cm 0cm 0.cm 0cm,clip=true]{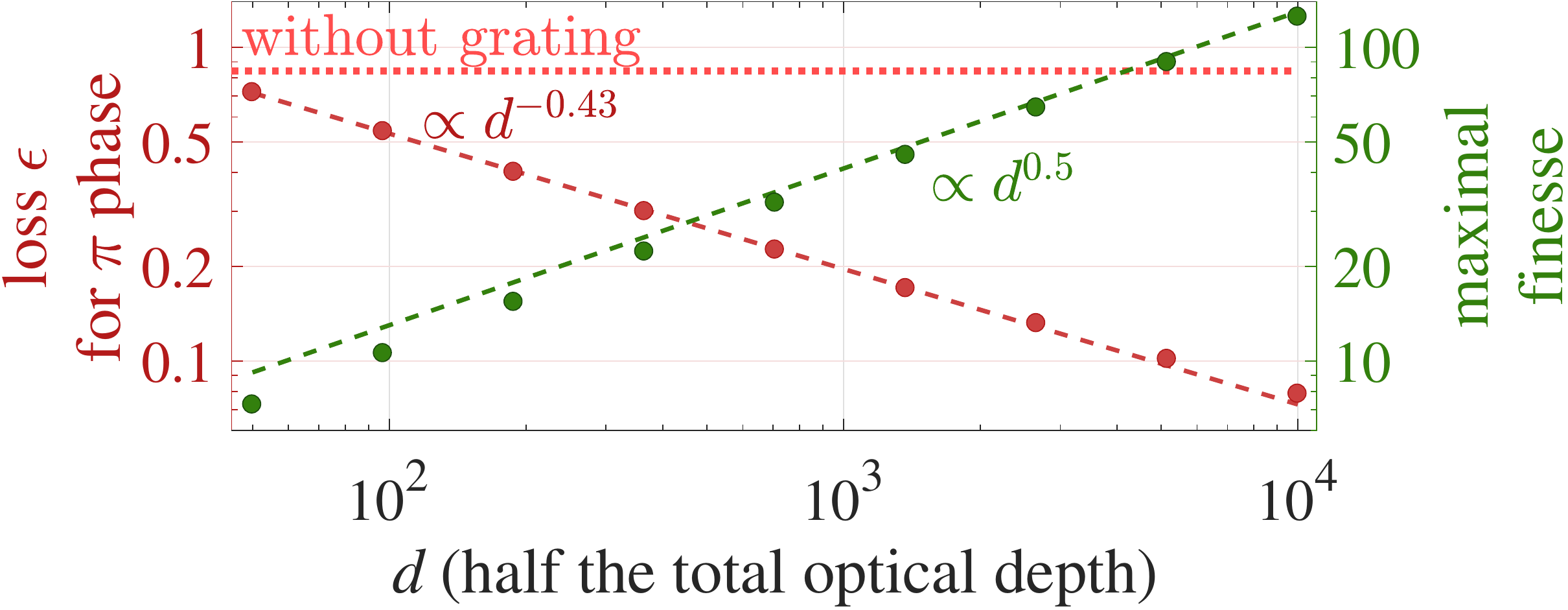}
\caption{Scaling of performances with optical depth. Green: numerical maximization of the induced cavity finesse (dots) compared to the analytic result $\FF\approx 1.3\sqrt{d}$ (line). Red: numerical minimization of the loss for a phase gate with $2\db=4\pi$ (dots) and a power-law fit (line). For example, for $d=10^4$, the optimization finds $\epsilon=0.08$ at $x= 6.0\times 10^{-4},~y=5.2\times 10^{-4}$. The dashed lines in Fig.~\ref{fig2}(b,c) are plotted for $y=5.2\times 10^{-4}$ and scanning $x$. Without the induced grating, the loss is very high ($\epsilon\approx 1$; horizontal orange line). See Ref.~\cite{SM} for more details on the optimized parameters.
}\label{fig4}
\end{figure}

We focus on the first resonance $\lambda L=\pi(1+i\alpha)$. Assuming a frequency modulation well within the EIT line $x,y\ll 1$, the absorption is given by $\alpha= d^2x^3/\pi^2$. With this and Eq.~(\ref{eqn:sagTnB}), in the large $d$ limit,
the finesse is insensitive to $\Delta/\Gamma$ and becomes $\FF=(1+\sqrt{T})\sqrt{(1-T)d}$ (see Ref.~\cite{SM} for details). The maximal finesse $\FF\approx 1.3 \sqrt{d}$ is obtained for $T=1/4$, as shown in Fig.~\ref{fig4}.

The loss $T < 1$ is of course the unavoidable downside of using atomic resonances, and we desire to maximize $T$ and $\FF$ simultaneously. For $1-T \ll 1$, the tradeoff arising from $\FF\propto \sqrt{1-T}$ can be heuristically estimated by adding $1-T$ to the loss $\epsilon$ in Eq.~(\ref{eqn:ring_error}) and minimizing $\epsilon$ with respect to $\FF$. The result is $\FF\approx\sqrt[3]{8\pi d /\db}$, which scales as $\sqrt[3]{d}$ rather than $\sqrt{d}$ \cite{SM}.

We find by numerical optimization that the exact performance of the scheme is slightly better than the above estimation.
In the numerics, we describe the blockade effect around the stored Rydberg excitation using the susceptibility $\chib$ at $|z-L/2|<\rb$. Outside the blockade volume, we use Eq.~(\ref{eqn:gratingProp}), so that the medium transmission is described by
\begin{equation} \label{eqn:DefectMat}
\Evec(0)=\textbf{F}_{\frac{L}{2}-\rb}
\begin{pmatrix}
e^{-ik\rb \chib} & 0  \\ 0 & e^{ik\rb \chib}
\end{pmatrix}
\textbf{F}_{\frac{L}{2}-\rb} \Evec(L).
\end{equation}
We calculate the Sagnac output $i u_j=r_j-t_j$, substitute into Eq.~(\ref{eqn:phi_eps}),
and minimize the loss $\epsilon$ while requiring $\phi=\pi$. As shown in Fig.~\ref{fig4}, the optimization finds that $\epsilon$ scales as $\sim d ^{-0.43}$. We conclude that optically-induced cavities can enhance the performance of photonic quantum gates.

We have examined a specific extension of EIT, utilizing far-detuned dressing. An alternative extension is a dual-V configuration rendering ``stationary light'' \cite{Zimmer2006,Andre2005e,Bajcsy2003a,Sorenson2017,SorensonArxiv2016Sagnac}. Here counter-propagating control fields couple the two propagation directions of the probe fields via resonant four-wave mixing.
We have repeated our analysis for this scheme. As shown by Hafezi \emph{et al.}~\cite{Hafezi2012}, this scheme affords a higher maximal finesse $\FF \propto d$, but our optimization yields an overall scaling of $1/\epsilon$ slower than $\sqrt{d}$ when accounting for the reduced transmission $T<1$, as also reported in Ref.~\cite{SorensonArxiv2016Sagnac}.

\emph{Conclusions.}---We showed that optical modulation of a finite medium can form an effective cavity that strengthens photon-photon interactions based on Rydberg blockade. The inadequacy of a symmetric (two-port) cavity for a conditional $\pi$-phase operation is solved with a Sagnac interferometer.
We benchmark the induced cavity by calculating the phase/loss relation in a Rydberg-based phase gate. This relation is described approximately by a circle, whose radius scales linearly with $\db\FF$. For the specific atomic system we consider, the finesse of the induced cavity scales roughly as $\FF\propto d^{0.4}$ (when optimized together with the overall transmission), and thus so is the effective enhancement of $\db$. By this we establish that the optical depth of the medium outside the blockade volume is a complementary resource to the limited $\db$. The essential ingredient of the scheme is the coupling between the counter-propagating modes. This coupling delays an incoming probe pulse via so-called `structural' slow light, as opposed to the standard delay due to EIT, pertaining to `material' slow light \cite{Boyd2011}. While the latter maintains the amplitude of the incoming probe field, the former increases it in the medium.

Optically-induced bandgaps were demonstrated experimentally using several atomic level schemes \cite{Bajcsy2003a,KatzHowell2013,Lin2009}. The proposed induced cavities are realizable with the optical depths of $10^3-10^5$ obtained with either cold \cite{Sparkes2013a,Blatt2014} or hot \cite{Walmsley2015OD} atoms. As a general concept, induced cavities could be employed in other systems, where actual cavities are impractical or for switchable functionality of photonic devices.

\bibliographystyle{apsrev4-1}

\begin{thebibliography}{68}%
\makeatletter
\providecommand \@ifxundefined [1]{%
 \@ifx{#1\undefined}
}%
\providecommand \@ifnum [1]{%
 \ifnum #1\expandafter \@firstoftwo
 \else \expandafter \@secondoftwo
 \fi
}%
\providecommand \@ifx [1]{%
 \ifx #1\expandafter \@firstoftwo
 \else \expandafter \@secondoftwo
 \fi
}%
\providecommand \natexlab [1]{#1}%
\providecommand \enquote  [1]{``#1''}%
\providecommand \bibnamefont  [1]{#1}%
\providecommand \bibfnamefont [1]{#1}%
\providecommand \citenamefont [1]{#1}%
\providecommand \href@noop [0]{\@secondoftwo}%
\providecommand \href [0]{\begingroup \@sanitize@url \@href}%
\providecommand \@href[1]{\@@startlink{#1}\@@href}%
\providecommand \@@href[1]{\endgroup#1\@@endlink}%
\providecommand \@sanitize@url [0]{\catcode `\\12\catcode `\$12\catcode
  `\&12\catcode `\#12\catcode `\^12\catcode `\_12\catcode `\%12\relax}%
\providecommand \@@startlink[1]{}%
\providecommand \@@endlink[0]{}%
\providecommand \url  [0]{\begingroup\@sanitize@url \@url }%
\providecommand \@url [1]{\endgroup\@href {#1}{\urlprefix }}%
\providecommand \urlprefix  [0]{URL }%
\providecommand \Eprint [0]{\href }%
\providecommand \doibase [0]{http://dx.doi.org/}%
\providecommand \selectlanguage [0]{\@gobble}%
\providecommand \bibinfo  [0]{\@secondoftwo}%
\providecommand \bibfield  [0]{\@secondoftwo}%
\providecommand \translation [1]{[#1]}%
\providecommand \BibitemOpen [0]{}%
\providecommand \bibitemStop [0]{}%
\providecommand \bibitemNoStop [0]{.\EOS\space}%
\providecommand \EOS [0]{\spacefactor3000\relax}%
\providecommand \BibitemShut  [1]{\csname bibitem#1\endcsname}%
\let\auto@bib@innerbib\@empty
\bibitem [{\citenamefont {Carusotto}\ and\ \citenamefont
  {Ciuti}(2013)}]{Carusotto2013}%
  \BibitemOpen
  \bibfield  {author} {\bibinfo {author} {\bibfnamefont {I.}~\bibnamefont
  {Carusotto}}\ and\ \bibinfo {author} {\bibfnamefont {C.}~\bibnamefont
  {Ciuti}},\ }\href {\doibase 10.1103/RevModPhys.85.299} {\bibfield  {journal}
  {\bibinfo  {journal} {Rev. Mod. Phys.}\ }\textbf {\bibinfo {volume} {85}},\
  \bibinfo {pages} {299} (\bibinfo {year} {2013})}\BibitemShut {NoStop}%
\bibitem [{\citenamefont {Otterbach}\ \emph {et~al.}(2013)\citenamefont
  {Otterbach}, \citenamefont {Moos}, \citenamefont {Muth},\ and\ \citenamefont
  {Fleischhauer}}]{Otterbach2013}%
  \BibitemOpen
  \bibfield  {author} {\bibinfo {author} {\bibfnamefont {J.}~\bibnamefont
  {Otterbach}}, \bibinfo {author} {\bibfnamefont {M.}~\bibnamefont {Moos}},
  \bibinfo {author} {\bibfnamefont {D.}~\bibnamefont {Muth}}, \ and\ \bibinfo
  {author} {\bibfnamefont {M.}~\bibnamefont {Fleischhauer}},\ }\href {\doibase
  10.1103/PhysRevLett.111.113001} {\bibfield  {journal} {\bibinfo  {journal}
  {Phys. Rev. Lett.}\ }\textbf {\bibinfo {volume} {111}},\ \bibinfo {pages}
  {113001} (\bibinfo {year} {2013})}\BibitemShut {NoStop}%
\bibitem [{\citenamefont {Shahmoon}\ \emph {et~al.}(2016)\citenamefont
  {Shahmoon}, \citenamefont {Gri{\v{s}}ins}, \citenamefont {Stimming},
  \citenamefont {Mazets},\ and\ \citenamefont {Kurizki}}]{Shahmoon2014}%
  \BibitemOpen
  \bibfield  {author} {\bibinfo {author} {\bibfnamefont {E.}~\bibnamefont
  {Shahmoon}}, \bibinfo {author} {\bibfnamefont {P.}~\bibnamefont
  {Gri{\v{s}}ins}}, \bibinfo {author} {\bibfnamefont {H.~P.}\ \bibnamefont
  {Stimming}}, \bibinfo {author} {\bibfnamefont {I.}~\bibnamefont {Mazets}}, \
  and\ \bibinfo {author} {\bibfnamefont {G.}~\bibnamefont {Kurizki}},\ }\href
  {\doibase 10.1364/OPTICA.3.000725} {\bibfield  {journal} {\bibinfo  {journal}
  {Optica}\ }\textbf {\bibinfo {volume} {3}},\ \bibinfo {pages} {725} (\bibinfo
  {year} {2016})}\BibitemShut {NoStop}%
\bibitem [{\citenamefont {Maghrebi}\ \emph {et~al.}(2015)\citenamefont
  {Maghrebi}, \citenamefont {Gullans}, \citenamefont {Bienias}, \citenamefont
  {Choi}, \citenamefont {Martin}, \citenamefont {Firstenberg}, \citenamefont
  {Lukin}, \citenamefont {B{\"{u}}chler},\ and\ \citenamefont
  {Gorshkov}}]{Gorshkov2015}%
  \BibitemOpen
  \bibfield  {author} {\bibinfo {author} {\bibfnamefont {M.~F.}\ \bibnamefont
  {Maghrebi}}, \bibinfo {author} {\bibfnamefont {M.~J.}\ \bibnamefont
  {Gullans}}, \bibinfo {author} {\bibfnamefont {P.}~\bibnamefont {Bienias}},
  \bibinfo {author} {\bibfnamefont {S.}~\bibnamefont {Choi}}, \bibinfo {author}
  {\bibfnamefont {I.}~\bibnamefont {Martin}}, \bibinfo {author} {\bibfnamefont
  {O.}~\bibnamefont {Firstenberg}}, \bibinfo {author} {\bibfnamefont {M.~D.}\
  \bibnamefont {Lukin}}, \bibinfo {author} {\bibfnamefont {H.~P.}\ \bibnamefont
  {B{\"{u}}chler}}, \ and\ \bibinfo {author} {\bibfnamefont {A.~V.}\
  \bibnamefont {Gorshkov}},\ }\href {\doibase 10.1103/PhysRevLett.115.123601}
  {\bibfield  {journal} {\bibinfo  {journal} {Phys. Rev. Lett.}\ }\textbf
  {\bibinfo {volume} {115}},\ \bibinfo {pages} {123601} (\bibinfo {year}
  {2015})}\BibitemShut {NoStop}%
\bibitem [{\citenamefont {Firstenberg}\ \emph {et~al.}(2013)\citenamefont
  {Firstenberg}, \citenamefont {Peyronel}, \citenamefont {Liang}, \citenamefont
  {Gorshkov}, \citenamefont {Lukin},\ and\ \citenamefont
  {Vuleti{\'{c}}}}]{Firstenberg2013AttrPhot}%
  \BibitemOpen
  \bibfield  {author} {\bibinfo {author} {\bibfnamefont {O.}~\bibnamefont
  {Firstenberg}}, \bibinfo {author} {\bibfnamefont {T.}~\bibnamefont
  {Peyronel}}, \bibinfo {author} {\bibfnamefont {Q.-Y.}\ \bibnamefont {Liang}},
  \bibinfo {author} {\bibfnamefont {A.~V.}\ \bibnamefont {Gorshkov}}, \bibinfo
  {author} {\bibfnamefont {M.~D.}\ \bibnamefont {Lukin}}, \ and\ \bibinfo
  {author} {\bibfnamefont {V.}~\bibnamefont {Vuleti{\'{c}}}},\ }\href {\doibase
  10.1038/nature12512} {\bibfield  {journal} {\bibinfo  {journal} {Nature}\
  }\textbf {\bibinfo {volume} {502}},\ \bibinfo {pages} {71} (\bibinfo {year}
  {2013})}\BibitemShut {NoStop}%
\bibitem [{\citenamefont {O'Brien}(2008)}]{OBrien2007}%
  \BibitemOpen
  \bibfield  {author} {\bibinfo {author} {\bibfnamefont {J.~L.}\ \bibnamefont
  {O'Brien}},\ }\href {\doibase 10.1126/science.1142892} {\bibfield  {journal}
  {\bibinfo  {journal} {Science}\ }\textbf {\bibinfo {volume} {318}},\ \bibinfo
  {pages} {1567} (\bibinfo {year} {2008})}\BibitemShut {NoStop}%
\bibitem [{\citenamefont {O'Brien}\ \emph {et~al.}(2010)\citenamefont
  {O'Brien}, \citenamefont {Furusawa},\ and\ \citenamefont
  {Vu{\v{c}}kovi{\'{c}}}}]{OBrien2009}%
  \BibitemOpen
  \bibfield  {author} {\bibinfo {author} {\bibfnamefont {J.~L.}\ \bibnamefont
  {O'Brien}}, \bibinfo {author} {\bibfnamefont {A.}~\bibnamefont {Furusawa}}, \
  and\ \bibinfo {author} {\bibfnamefont {J.}~\bibnamefont
  {Vu{\v{c}}kovi{\'{c}}}},\ }\href {\doibase 10.1038/nphoton.2009.229}
  {\bibfield  {journal} {\bibinfo  {journal} {Nat. Photonics}\ }\textbf
  {\bibinfo {volume} {3}},\ \bibinfo {pages} {687} (\bibinfo {year}
  {2010})}\BibitemShut {NoStop}%
\bibitem [{\citenamefont {Kimble}(2008)}]{Kimble2008}%
  \BibitemOpen
  \bibfield  {author} {\bibinfo {author} {\bibfnamefont {H.~J.}\ \bibnamefont
  {Kimble}},\ }\href {\doibase 10.1038/nature07127} {\bibfield  {journal}
  {\bibinfo  {journal} {Nature}\ }\textbf {\bibinfo {volume} {453}},\ \bibinfo
  {pages} {1023} (\bibinfo {year} {2008})}\BibitemShut {NoStop}%
\bibitem [{\citenamefont {Chang}\ \emph {et~al.}(2014)\citenamefont {Chang},
  \citenamefont {Vuleti{\'{c}}},\ and\ \citenamefont {Lukin}}]{Chang2014}%
  \BibitemOpen
  \bibfield  {author} {\bibinfo {author} {\bibfnamefont {D.~E.}\ \bibnamefont
  {Chang}}, \bibinfo {author} {\bibfnamefont {V.}~\bibnamefont
  {Vuleti{\'{c}}}}, \ and\ \bibinfo {author} {\bibfnamefont {M.~D.}\
  \bibnamefont {Lukin}},\ }\href {\doibase 10.1038/nphoton.2014.192} {\bibfield
   {journal} {\bibinfo  {journal} {Nat. Photonics}\ }\textbf {\bibinfo {volume}
  {8}},\ \bibinfo {pages} {685} (\bibinfo {year} {2014})}\BibitemShut {NoStop}%
\bibitem [{\citenamefont {Roy}\ \emph {et~al.}(2017)\citenamefont {Roy},
  \citenamefont {Wilson},\ and\ \citenamefont
  {Firstenberg}}]{FirstenbergRMP2017}%
  \BibitemOpen
  \bibfield  {author} {\bibinfo {author} {\bibfnamefont {D.}~\bibnamefont
  {Roy}}, \bibinfo {author} {\bibfnamefont {C.~M.}\ \bibnamefont {Wilson}}, \
  and\ \bibinfo {author} {\bibfnamefont {O.}~\bibnamefont {Firstenberg}},\
  }\href {\doibase 10.1103/RevModPhys.89.021001} {\bibfield  {journal}
  {\bibinfo  {journal} {Rev. Mod. Phys.}\ }\textbf {\bibinfo {volume} {89}},\
  \bibinfo {pages} {021001} (\bibinfo {year} {2017})}\BibitemShut {NoStop}%
\bibitem [{\citenamefont {Tanji-Suzuki}\ \emph {et~al.}(2011)\citenamefont
  {Tanji-Suzuki}, \citenamefont {Leroux}, \citenamefont {Schleier-Smith},
  \citenamefont {Cetina}, \citenamefont {Grier}, \citenamefont {Simon},\ and\
  \citenamefont {Vuleti{\'{c}}}}]{Vuletic2011}%
  \BibitemOpen
  \bibfield  {author} {\bibinfo {author} {\bibfnamefont {H.}~\bibnamefont
  {Tanji-Suzuki}}, \bibinfo {author} {\bibfnamefont {I.~D.}\ \bibnamefont
  {Leroux}}, \bibinfo {author} {\bibfnamefont {M.~H.}\ \bibnamefont
  {Schleier-Smith}}, \bibinfo {author} {\bibfnamefont {M.}~\bibnamefont
  {Cetina}}, \bibinfo {author} {\bibfnamefont {A.~T.}\ \bibnamefont {Grier}},
  \bibinfo {author} {\bibfnamefont {J.}~\bibnamefont {Simon}}, \ and\ \bibinfo
  {author} {\bibfnamefont {V.}~\bibnamefont {Vuleti{\'{c}}}},\ }\href {\doibase
  10.1016/B978-0-12-385508-4.00004-8} {\bibfield  {journal} {\bibinfo
  {journal} {Adv. At. Mol. Opt. Phys.}\ }\textbf {\bibinfo {volume} {60}},\
  \bibinfo {pages} {201} (\bibinfo {year} {2011})}\BibitemShut {NoStop}%
\bibitem [{\citenamefont {Duan}\ and\ \citenamefont
  {Kimble}(2004)}]{Kimble2004}%
  \BibitemOpen
  \bibfield  {author} {\bibinfo {author} {\bibfnamefont {L.~M.}\ \bibnamefont
  {Duan}}\ and\ \bibinfo {author} {\bibfnamefont {H.~J.}\ \bibnamefont
  {Kimble}},\ }\href {\doibase 10.1103/PhysRevLett.92.127902} {\bibfield
  {journal} {\bibinfo  {journal} {Phys. Rev. Lett.}\ }\textbf {\bibinfo
  {volume} {92}},\ \bibinfo {pages} {127902} (\bibinfo {year}
  {2004})}\BibitemShut {NoStop}%
\bibitem [{\citenamefont {Hacker}\ \emph {et~al.}(2016)\citenamefont {Hacker},
  \citenamefont {Welte}, \citenamefont {Rempe},\ and\ \citenamefont
  {Ritter}}]{Rempe2016Gate}%
  \BibitemOpen
  \bibfield  {author} {\bibinfo {author} {\bibfnamefont {B.}~\bibnamefont
  {Hacker}}, \bibinfo {author} {\bibfnamefont {S.}~\bibnamefont {Welte}},
  \bibinfo {author} {\bibfnamefont {G.}~\bibnamefont {Rempe}}, \ and\ \bibinfo
  {author} {\bibfnamefont {S.}~\bibnamefont {Ritter}},\ }\href {\doibase
  10.1038/nature18592} {\bibfield  {journal} {\bibinfo  {journal} {Nature}\
  }\textbf {\bibinfo {volume} {536}},\ \bibinfo {pages} {1} (\bibinfo {year}
  {2016})}\BibitemShut {NoStop}%
\bibitem [{\citenamefont {Tiecke}\ \emph {et~al.}(2014)\citenamefont {Tiecke},
  \citenamefont {Thompson}, \citenamefont {de~Leon}, \citenamefont {Liu},
  \citenamefont {Vuleti{\'{c}}},\ and\ \citenamefont {Lukin}}]{Lukin2014PC}%
  \BibitemOpen
  \bibfield  {author} {\bibinfo {author} {\bibfnamefont {T.~G.}\ \bibnamefont
  {Tiecke}}, \bibinfo {author} {\bibfnamefont {J.~D.}\ \bibnamefont
  {Thompson}}, \bibinfo {author} {\bibfnamefont {N.~P.}\ \bibnamefont
  {de~Leon}}, \bibinfo {author} {\bibfnamefont {L.~R.}\ \bibnamefont {Liu}},
  \bibinfo {author} {\bibfnamefont {V.}~\bibnamefont {Vuleti{\'{c}}}}, \ and\
  \bibinfo {author} {\bibfnamefont {M.~D.}\ \bibnamefont {Lukin}},\ }\href
  {\doibase 10.1038/nature13188} {\bibfield  {journal} {\bibinfo  {journal}
  {Nature}\ }\textbf {\bibinfo {volume} {508}},\ \bibinfo {pages} {241}
  (\bibinfo {year} {2014})}\BibitemShut {NoStop}%
\bibitem [{\citenamefont {Volz}\ \emph {et~al.}(2014)\citenamefont {Volz},
  \citenamefont {Scheucher}, \citenamefont {Junge},\ and\ \citenamefont
  {Rauschenbeutel}}]{Rauschen2014Pi}%
  \BibitemOpen
  \bibfield  {author} {\bibinfo {author} {\bibfnamefont {J.}~\bibnamefont
  {Volz}}, \bibinfo {author} {\bibfnamefont {M.}~\bibnamefont {Scheucher}},
  \bibinfo {author} {\bibfnamefont {C.}~\bibnamefont {Junge}}, \ and\ \bibinfo
  {author} {\bibfnamefont {A.}~\bibnamefont {Rauschenbeutel}},\ }\href
  {\doibase 10.1038/nphoton.2014.253} {\bibfield  {journal} {\bibinfo
  {journal} {Nat. Photonics}\ }\textbf {\bibinfo {volume} {8}},\ \bibinfo
  {pages} {965} (\bibinfo {year} {2014})}\BibitemShut {NoStop}%
\bibitem [{\citenamefont {Firstenberg}\ \emph {et~al.}(2016)\citenamefont
  {Firstenberg}, \citenamefont {Adams},\ and\ \citenamefont
  {Hofferberth}}]{FirstenbergReviewJPB2016}%
  \BibitemOpen
  \bibfield  {author} {\bibinfo {author} {\bibfnamefont {O.}~\bibnamefont
  {Firstenberg}}, \bibinfo {author} {\bibfnamefont {C.~S.}\ \bibnamefont
  {Adams}}, \ and\ \bibinfo {author} {\bibfnamefont {S.}~\bibnamefont
  {Hofferberth}},\ }\href {http://stacks.iop.org/0953-4075/49/i=15/a=152003}
  {\bibfield  {journal} {\bibinfo  {journal} {Journal of Physics B: Atomic,
  Molecular and Optical Physics}\ }\textbf {\bibinfo {volume} {49}},\ \bibinfo
  {pages} {152003} (\bibinfo {year} {2016})}\BibitemShut {NoStop}%
\bibitem [{\citenamefont {Peyronel}\ \emph
  {et~al.}(2012{\natexlab{a}})\citenamefont {Peyronel}, \citenamefont
  {Firstenberg}, \citenamefont {Liang}, \citenamefont {Hofferberth},
  \citenamefont {Gorshkov}, \citenamefont {Pohl}, \citenamefont {Lukin},\ and\
  \citenamefont {Vuleti{\'{c}}}}]{Peyronel2012}%
  \BibitemOpen
  \bibfield  {author} {\bibinfo {author} {\bibfnamefont {T.}~\bibnamefont
  {Peyronel}}, \bibinfo {author} {\bibfnamefont {O.}~\bibnamefont
  {Firstenberg}}, \bibinfo {author} {\bibfnamefont {Q.-Y.}\ \bibnamefont
  {Liang}}, \bibinfo {author} {\bibfnamefont {S.}~\bibnamefont {Hofferberth}},
  \bibinfo {author} {\bibfnamefont {A.~V.}\ \bibnamefont {Gorshkov}}, \bibinfo
  {author} {\bibfnamefont {T.}~\bibnamefont {Pohl}}, \bibinfo {author}
  {\bibfnamefont {M.~D.}\ \bibnamefont {Lukin}}, \ and\ \bibinfo {author}
  {\bibfnamefont {V.}~\bibnamefont {Vuleti{\'{c}}}},\ }\href {\doibase
  10.1038/nature11361} {\bibfield  {journal} {\bibinfo  {journal} {Nature}\
  }\textbf {\bibinfo {volume} {488}},\ \bibinfo {pages} {57} (\bibinfo {year}
  {2012}{\natexlab{a}})}\BibitemShut {NoStop}%
\bibitem [{\citenamefont {Tiarks}\ \emph {et~al.}(2016)\citenamefont {Tiarks},
  \citenamefont {Schmidt}, \citenamefont {Rempe},\ and\ \citenamefont
  {Du~rr}}]{Durr2016}%
  \BibitemOpen
  \bibfield  {author} {\bibinfo {author} {\bibfnamefont {D.}~\bibnamefont
  {Tiarks}}, \bibinfo {author} {\bibfnamefont {S.}~\bibnamefont {Schmidt}},
  \bibinfo {author} {\bibfnamefont {G.}~\bibnamefont {Rempe}}, \ and\ \bibinfo
  {author} {\bibfnamefont {S.}~\bibnamefont {Du~rr}},\ }\href {\doibase
  10.1126/sciadv.1600036} {\bibfield  {journal} {\bibinfo  {journal} {Sci.
  Adv.}\ }\textbf {\bibinfo {volume} {2}},\ \bibinfo {pages} {e1600036}
  (\bibinfo {year} {2016})}\BibitemShut {NoStop}%
\bibitem [{\citenamefont {Tresp}\ \emph {et~al.}(2016)\citenamefont {Tresp},
  \citenamefont {Zimmer}, \citenamefont {Mirgorodskiy}, \citenamefont
  {Gorniaczyk}, \citenamefont {Paris-Mandoki},\ and\ \citenamefont
  {Hofferberth}}]{Tresp2016Absorber}%
  \BibitemOpen
  \bibfield  {author} {\bibinfo {author} {\bibfnamefont {C.}~\bibnamefont
  {Tresp}}, \bibinfo {author} {\bibfnamefont {C.}~\bibnamefont {Zimmer}},
  \bibinfo {author} {\bibfnamefont {I.}~\bibnamefont {Mirgorodskiy}}, \bibinfo
  {author} {\bibfnamefont {H.}~\bibnamefont {Gorniaczyk}}, \bibinfo {author}
  {\bibfnamefont {A.}~\bibnamefont {Paris-Mandoki}}, \ and\ \bibinfo {author}
  {\bibfnamefont {S.}~\bibnamefont {Hofferberth}},\ }\href {\doibase
  10.1103/PhysRevLett.117.223001} {\bibfield  {journal} {\bibinfo  {journal}
  {Phys. Rev. Lett.}\ }\textbf {\bibinfo {volume} {117}},\ \bibinfo {pages}
  {223001} (\bibinfo {year} {2016})}\BibitemShut {NoStop}%
\bibitem [{\citenamefont {Pritchard}\ \emph {et~al.}(2013)\citenamefont
  {Pritchard}, \citenamefont {Weatherill},\ and\ \citenamefont
  {Adams}}]{AdamsReview2013}%
  \BibitemOpen
  \bibfield  {author} {\bibinfo {author} {\bibfnamefont {J.~D.}\ \bibnamefont
  {Pritchard}}, \bibinfo {author} {\bibfnamefont {K.~J.}\ \bibnamefont
  {Weatherill}}, \ and\ \bibinfo {author} {\bibfnamefont {C.~S.}\ \bibnamefont
  {Adams}},\ }\href {\doibase 10.1142/9789814440400_0008} {\bibfield  {journal}
  {\bibinfo  {journal} {Annual Review of Cold Atoms and Molecules}\ }\textbf
  {\bibinfo {volume} {1}},\ \bibinfo {pages} {301} (\bibinfo {year}
  {2013})}\BibitemShut {NoStop}%
\bibitem [{\citenamefont {Lukin}\ \emph {et~al.}(2000)\citenamefont {Lukin},
  \citenamefont {Fleischhauer}, \citenamefont {Cote}, \citenamefont {Duan},
  \citenamefont {Jaksch}, \citenamefont {Cirac},\ and\ \citenamefont
  {Zoller}}]{Lukin2001}%
  \BibitemOpen
  \bibfield  {author} {\bibinfo {author} {\bibfnamefont {M.~D.}\ \bibnamefont
  {Lukin}}, \bibinfo {author} {\bibfnamefont {M.}~\bibnamefont {Fleischhauer}},
  \bibinfo {author} {\bibfnamefont {R.}~\bibnamefont {Cote}}, \bibinfo {author}
  {\bibfnamefont {L.~M.}\ \bibnamefont {Duan}}, \bibinfo {author}
  {\bibfnamefont {D.}~\bibnamefont {Jaksch}}, \bibinfo {author} {\bibfnamefont
  {J.~I.}\ \bibnamefont {Cirac}}, \ and\ \bibinfo {author} {\bibfnamefont
  {P.}~\bibnamefont {Zoller}},\ }\href {\doibase 10.1103/PhysRevLett.87.037901}
  {\bibfield  {journal} {\bibinfo  {journal} {Phys. Rev. Lett.}\ }\textbf
  {\bibinfo {volume} {87}},\ \bibinfo {pages} {037901} (\bibinfo {year}
  {2000})}\BibitemShut {NoStop}%
\bibitem [{\citenamefont {Saffman}\ \emph {et~al.}(2010)\citenamefont
  {Saffman}, \citenamefont {Walker},\ and\ \citenamefont
  {M{\o}lmer}}]{Saffman2010RydbergRev}%
  \BibitemOpen
  \bibfield  {author} {\bibinfo {author} {\bibfnamefont {M.}~\bibnamefont
  {Saffman}}, \bibinfo {author} {\bibfnamefont {T.~G.}\ \bibnamefont {Walker}},
  \ and\ \bibinfo {author} {\bibfnamefont {K.}~\bibnamefont {M{\o}lmer}},\
  }\href {\doibase 10.1103/RevModPhys.82.2313} {\bibfield  {journal} {\bibinfo
  {journal} {Rev. Mod. Phys.}\ }\textbf {\bibinfo {volume} {82}},\ \bibinfo
  {pages} {2313} (\bibinfo {year} {2010})}\BibitemShut {NoStop}%
\bibitem [{\citenamefont {Pritchard}\ \emph {et~al.}(2010)\citenamefont
  {Pritchard}, \citenamefont {Maxwell}, \citenamefont {Gauguet}, \citenamefont
  {Weatherill}, \citenamefont {Jones},\ and\ \citenamefont
  {Adams}}]{PritchardPRL2010}%
  \BibitemOpen
  \bibfield  {author} {\bibinfo {author} {\bibfnamefont {J.~D.}\ \bibnamefont
  {Pritchard}}, \bibinfo {author} {\bibfnamefont {D.}~\bibnamefont {Maxwell}},
  \bibinfo {author} {\bibfnamefont {A.}~\bibnamefont {Gauguet}}, \bibinfo
  {author} {\bibfnamefont {K.~J.}\ \bibnamefont {Weatherill}}, \bibinfo
  {author} {\bibfnamefont {M.~P.~A.}\ \bibnamefont {Jones}}, \ and\ \bibinfo
  {author} {\bibfnamefont {C.~S.}\ \bibnamefont {Adams}},\ }\href {\doibase
  10.1103/PhysRevLett.105.193603} {\bibfield  {journal} {\bibinfo  {journal}
  {Phys. Rev. Lett.}\ }\textbf {\bibinfo {volume} {105}},\ \bibinfo {pages}
  {193603} (\bibinfo {year} {2010})}\BibitemShut {NoStop}%
\bibitem [{\citenamefont {Vuletic}(2006)}]{vuletic2006superatom}%
  \BibitemOpen
  \bibfield  {author} {\bibinfo {author} {\bibfnamefont {V.}~\bibnamefont
  {Vuletic}},\ }\href {\doibase 10.1038/nphys469} {\bibfield  {journal}
  {\bibinfo  {journal} {Nat. Phys.}\ }\textbf {\bibinfo {volume} {2}},\
  \bibinfo {pages} {801} (\bibinfo {year} {2006})}\BibitemShut {NoStop}%
\bibitem [{\citenamefont {Johnson}\ \emph {et~al.}(2008)\citenamefont
  {Johnson}, \citenamefont {Urban}, \citenamefont {Henage}, \citenamefont
  {Isenhower}, \citenamefont {Yavuz}, \citenamefont {Walker},\ and\
  \citenamefont {Saffman}}]{Saffman2008}%
  \BibitemOpen
  \bibfield  {author} {\bibinfo {author} {\bibfnamefont {T.~A.}\ \bibnamefont
  {Johnson}}, \bibinfo {author} {\bibfnamefont {E.}~\bibnamefont {Urban}},
  \bibinfo {author} {\bibfnamefont {T.}~\bibnamefont {Henage}}, \bibinfo
  {author} {\bibfnamefont {L.}~\bibnamefont {Isenhower}}, \bibinfo {author}
  {\bibfnamefont {D.~D.}\ \bibnamefont {Yavuz}}, \bibinfo {author}
  {\bibfnamefont {T.~G.}\ \bibnamefont {Walker}}, \ and\ \bibinfo {author}
  {\bibfnamefont {M.}~\bibnamefont {Saffman}},\ }\href {\doibase
  10.1103/PhysRevLett.100.113003} {\bibfield  {journal} {\bibinfo  {journal}
  {Phys. Rev. Lett.}\ }\textbf {\bibinfo {volume} {100}},\ \bibinfo {pages} {2}
  (\bibinfo {year} {2008})}\BibitemShut {NoStop}%
\bibitem [{\citenamefont {Dudin}\ \emph {et~al.}(2012)\citenamefont {Dudin},
  \citenamefont {Li}, \citenamefont {Bariani},\ and\ \citenamefont
  {Kuzmich}}]{Dudin2012}%
  \BibitemOpen
  \bibfield  {author} {\bibinfo {author} {\bibfnamefont {Y.~O.}\ \bibnamefont
  {Dudin}}, \bibinfo {author} {\bibfnamefont {L.}~\bibnamefont {Li}}, \bibinfo
  {author} {\bibfnamefont {F.}~\bibnamefont {Bariani}}, \ and\ \bibinfo
  {author} {\bibfnamefont {A.}~\bibnamefont {Kuzmich}},\ }\href {\doibase
  10.1038/nphys2413} {\bibfield  {journal} {\bibinfo  {journal} {Nat. Phys.}\
  }\textbf {\bibinfo {volume} {8}},\ \bibinfo {pages} {790} (\bibinfo {year}
  {2012})}\BibitemShut {NoStop}%
\bibitem [{\citenamefont {Peyronel}\ \emph
  {et~al.}(2012{\natexlab{b}})\citenamefont {Peyronel}, \citenamefont
  {Firstenberg}, \citenamefont {Liang}, \citenamefont {Hofferberth},
  \citenamefont {Gorshkov}, \citenamefont {Pohl}, \citenamefont {Lukin},\ and\
  \citenamefont {Vuletic}}]{PeyronelNature2012}%
  \BibitemOpen
  \bibfield  {author} {\bibinfo {author} {\bibfnamefont {T.}~\bibnamefont
  {Peyronel}}, \bibinfo {author} {\bibfnamefont {O.}~\bibnamefont
  {Firstenberg}}, \bibinfo {author} {\bibfnamefont {Q.-Y.}\ \bibnamefont
  {Liang}}, \bibinfo {author} {\bibfnamefont {S.}~\bibnamefont {Hofferberth}},
  \bibinfo {author} {\bibfnamefont {A.~V.}\ \bibnamefont {Gorshkov}}, \bibinfo
  {author} {\bibfnamefont {T.}~\bibnamefont {Pohl}}, \bibinfo {author}
  {\bibfnamefont {M.~D.}\ \bibnamefont {Lukin}}, \ and\ \bibinfo {author}
  {\bibfnamefont {V.}~\bibnamefont {Vuletic}},\ }\href@noop {} {\bibfield
  {journal} {\bibinfo  {journal} {Nature (London)}\ }\textbf {\bibinfo {volume}
  {488}},\ \bibinfo {pages} {57} (\bibinfo {year}
  {2012}{\natexlab{b}})}\BibitemShut {NoStop}%
\bibitem [{\citenamefont {Gorshkov}\ \emph {et~al.}(2013)\citenamefont
  {Gorshkov}, \citenamefont {Nath},\ and\ \citenamefont {Pohl}}]{Gorshkov2013}%
  \BibitemOpen
  \bibfield  {author} {\bibinfo {author} {\bibfnamefont {A.~V.}\ \bibnamefont
  {Gorshkov}}, \bibinfo {author} {\bibfnamefont {R.}~\bibnamefont {Nath}}, \
  and\ \bibinfo {author} {\bibfnamefont {T.}~\bibnamefont {Pohl}},\ }\href
  {\doibase 10.1103/PhysRevLett.110.153601} {\bibfield  {journal} {\bibinfo
  {journal} {Phys. Rev. Lett.}\ }\textbf {\bibinfo {volume} {110}},\ \bibinfo
  {pages} {153601} (\bibinfo {year} {2013})}\BibitemShut {NoStop}%
\bibitem [{\citenamefont {Paredes-Barato}\ and\ \citenamefont
  {Adams}(2014)}]{Adams2014}%
  \BibitemOpen
  \bibfield  {author} {\bibinfo {author} {\bibfnamefont {D.}~\bibnamefont
  {Paredes-Barato}}\ and\ \bibinfo {author} {\bibfnamefont {C.~S.}\
  \bibnamefont {Adams}},\ }\href {\doibase 10.1103/PhysRevLett.112.040501}
  {\bibfield  {journal} {\bibinfo  {journal} {Phys. Rev. Lett.}\ }\textbf
  {\bibinfo {volume} {112}},\ \bibinfo {pages} {040501} (\bibinfo {year}
  {2014})}\BibitemShut {NoStop}%
\bibitem [{\citenamefont {Moos}\ \emph {et~al.}(2015)\citenamefont {Moos},
  \citenamefont {H\"oning}, \citenamefont {Unanyan},\ and\ \citenamefont
  {Fleischhauer}}]{Moos2015}%
  \BibitemOpen
  \bibfield  {author} {\bibinfo {author} {\bibfnamefont {M.}~\bibnamefont
  {Moos}}, \bibinfo {author} {\bibfnamefont {M.}~\bibnamefont {H\"oning}},
  \bibinfo {author} {\bibfnamefont {R.}~\bibnamefont {Unanyan}}, \ and\
  \bibinfo {author} {\bibfnamefont {M.}~\bibnamefont {Fleischhauer}},\ }\href
  {\doibase 10.1103/PhysRevA.92.053846} {\bibfield  {journal} {\bibinfo
  {journal} {Phys. Rev. A}\ }\textbf {\bibinfo {volume} {92}},\ \bibinfo
  {pages} {053846} (\bibinfo {year} {2015})}\BibitemShut {NoStop}%
\bibitem [{\citenamefont {Murray}\ \emph {et~al.}(2016)\citenamefont {Murray},
  \citenamefont {Gorshkov},\ and\ \citenamefont {Pohl}}]{murray2016many}%
  \BibitemOpen
  \bibfield  {author} {\bibinfo {author} {\bibfnamefont {C.~R.}\ \bibnamefont
  {Murray}}, \bibinfo {author} {\bibfnamefont {A.~V.}\ \bibnamefont
  {Gorshkov}}, \ and\ \bibinfo {author} {\bibfnamefont {T.}~\bibnamefont
  {Pohl}},\ }\href {http://arxiv.org/abs/1607.01984} {\bibfield  {journal}
  {\bibinfo  {journal} {New J. Phys.}\ }\textbf {\bibinfo {volume} {18}}
  (\bibinfo {year} {2016})}\BibitemShut {NoStop}%
\bibitem [{\citenamefont {Murray}\ and\ \citenamefont {Pohl}(2017)}]{Pohl2017}%
  \BibitemOpen
  \bibfield  {author} {\bibinfo {author} {\bibfnamefont {C.~R.}\ \bibnamefont
  {Murray}}\ and\ \bibinfo {author} {\bibfnamefont {T.}~\bibnamefont {Pohl}},\
  }\href {http://arxiv.org/abs/1702.03763} {\ ,\ \bibinfo {pages} {66}
  (\bibinfo {year} {2017})},\ \Eprint {http://arxiv.org/abs/1702.03763}
  {arXiv:1702.03763} \BibitemShut {NoStop}%
\bibitem [{\citenamefont {Gaj}\ \emph {et~al.}(2014)\citenamefont {Gaj},
  \citenamefont {Krupp}, \citenamefont {Balewski}, \citenamefont {L{\"{o}}w},
  \citenamefont {Hofferberth},\ and\ \citenamefont {Pfau}}]{Pfau2014Density}%
  \BibitemOpen
  \bibfield  {author} {\bibinfo {author} {\bibfnamefont {A.}~\bibnamefont
  {Gaj}}, \bibinfo {author} {\bibfnamefont {A.~T.}\ \bibnamefont {Krupp}},
  \bibinfo {author} {\bibfnamefont {J.~B.}\ \bibnamefont {Balewski}}, \bibinfo
  {author} {\bibfnamefont {R.}~\bibnamefont {L{\"{o}}w}}, \bibinfo {author}
  {\bibfnamefont {S.}~\bibnamefont {Hofferberth}}, \ and\ \bibinfo {author}
  {\bibfnamefont {T.}~\bibnamefont {Pfau}},\ }\href {\doibase
  10.1038/ncomms5546} {\bibfield  {journal} {\bibinfo  {journal} {Nat.
  Commun.}\ }\textbf {\bibinfo {volume} {5}},\ \bibinfo {pages} {4546}
  (\bibinfo {year} {2014})}\BibitemShut {NoStop}%
\bibitem [{\citenamefont {Baur}\ \emph {et~al.}(2014)\citenamefont {Baur},
  \citenamefont {Tiarks}, \citenamefont {Rempe},\ and\ \citenamefont
  {D\"{u}rr}}]{DurrPRL2014}%
  \BibitemOpen
  \bibfield  {author} {\bibinfo {author} {\bibfnamefont {S.}~\bibnamefont
  {Baur}}, \bibinfo {author} {\bibfnamefont {D.}~\bibnamefont {Tiarks}},
  \bibinfo {author} {\bibfnamefont {G.}~\bibnamefont {Rempe}}, \ and\ \bibinfo
  {author} {\bibfnamefont {S.}~\bibnamefont {D\"{u}rr}},\ }\href {\doibase
  10.1103/PhysRevLett.112.073901} {\bibfield  {journal} {\bibinfo  {journal}
  {Phys. Rev. Lett.}\ }\textbf {\bibinfo {volume} {112}},\ \bibinfo {pages}
  {073901} (\bibinfo {year} {2014})}\BibitemShut {NoStop}%
\bibitem [{\citenamefont {Bienias}\ and\ \citenamefont
  {B\"{u}chler}(2016)}]{Hanspeter2016Kerr}%
  \BibitemOpen
  \bibfield  {author} {\bibinfo {author} {\bibfnamefont {P.}~\bibnamefont
  {Bienias}}\ and\ \bibinfo {author} {\bibfnamefont {H.~P.}\ \bibnamefont
  {B\"{u}chler}},\ }\href {\doibase 10.1088/1367-2630/aa50c3} {\bibfield
  {journal} {\bibinfo  {journal} {New J. Phys.}\ }\textbf {\bibinfo {volume}
  {18}},\ \bibinfo {pages} {123026} (\bibinfo {year} {2016})}\BibitemShut
  {NoStop}%
\bibitem [{\citenamefont {Li}\ and\ \citenamefont
  {Lesanovsky}(2015)}]{LesanovskyPRA2015}%
  \BibitemOpen
  \bibfield  {author} {\bibinfo {author} {\bibfnamefont {W.}~\bibnamefont
  {Li}}\ and\ \bibinfo {author} {\bibfnamefont {I.}~\bibnamefont
  {Lesanovsky}},\ }\href {\doibase 10.1103/PhysRevA.92.043828} {\bibfield
  {journal} {\bibinfo  {journal} {Phys. Rev. A}\ }\textbf {\bibinfo {volume}
  {92}},\ \bibinfo {pages} {043828} (\bibinfo {year} {2015})}\BibitemShut
  {NoStop}%
\bibitem [{\citenamefont {Andr{\'{e}}}\ and\ \citenamefont
  {Lukin}(2002)}]{Andre2002e}%
  \BibitemOpen
  \bibfield  {author} {\bibinfo {author} {\bibfnamefont {A.}~\bibnamefont
  {Andr{\'{e}}}}\ and\ \bibinfo {author} {\bibfnamefont {M.~D.}\ \bibnamefont
  {Lukin}},\ }\href {\doibase 10.1103/PhysRevLett.89.143602} {\bibfield
  {journal} {\bibinfo  {journal} {Phys. Rev. Lett.}\ }\textbf {\bibinfo
  {volume} {89}},\ \bibinfo {pages} {143602} (\bibinfo {year}
  {2002})}\BibitemShut {NoStop}%
\bibitem [{\citenamefont {Andr{\'{e}}}\ \emph {et~al.}(2005)\citenamefont
  {Andr{\'{e}}}, \citenamefont {Bajcsy}, \citenamefont {Zibrov},\ and\
  \citenamefont {Lukin}}]{Andre2005e}%
  \BibitemOpen
  \bibfield  {author} {\bibinfo {author} {\bibfnamefont {A.}~\bibnamefont
  {Andr{\'{e}}}}, \bibinfo {author} {\bibfnamefont {M.}~\bibnamefont {Bajcsy}},
  \bibinfo {author} {\bibfnamefont {A.~S.}\ \bibnamefont {Zibrov}}, \ and\
  \bibinfo {author} {\bibfnamefont {M.~D.}\ \bibnamefont {Lukin}},\ }\href
  {\doibase 10.1103/PhysRevLett.94.063902} {\bibfield  {journal} {\bibinfo
  {journal} {Phys. Rev. Lett.}\ }\textbf {\bibinfo {volume} {94}},\ \bibinfo
  {pages} {063902} (\bibinfo {year} {2005})}\BibitemShut {NoStop}%
\bibitem [{\citenamefont {Hafezi}\ \emph {et~al.}(2012)\citenamefont {Hafezi},
  \citenamefont {Chang}, \citenamefont {Gritsev}, \citenamefont {Demler},\ and\
  \citenamefont {Lukin}}]{Hafezi2012}%
  \BibitemOpen
  \bibfield  {author} {\bibinfo {author} {\bibfnamefont {M.}~\bibnamefont
  {Hafezi}}, \bibinfo {author} {\bibfnamefont {D.~E.}\ \bibnamefont {Chang}},
  \bibinfo {author} {\bibfnamefont {V.}~\bibnamefont {Gritsev}}, \bibinfo
  {author} {\bibfnamefont {E.}~\bibnamefont {Demler}}, \ and\ \bibinfo {author}
  {\bibfnamefont {M.~D.}\ \bibnamefont {Lukin}},\ }\href {\doibase
  10.1103/PhysRevA.85.013822} {\bibfield  {journal} {\bibinfo  {journal} {Phys.
  Rev. A}\ }\textbf {\bibinfo {volume} {85}},\ \bibinfo {pages} {013822}
  (\bibinfo {year} {2012})}\BibitemShut {NoStop}%
\bibitem [{\citenamefont {Parigi}\ \emph {et~al.}(2013)\citenamefont {Parigi},
  \citenamefont {Bimbard}, \citenamefont {Stanojevic}, \citenamefont
  {Hilliard}, \citenamefont {Nogrette}, \citenamefont {Tualle-Brouri},
  \citenamefont {Ourjoumtsev},\ and\ \citenamefont {Grangier}}]{Grangier2012}%
  \BibitemOpen
  \bibfield  {author} {\bibinfo {author} {\bibfnamefont {V.}~\bibnamefont
  {Parigi}}, \bibinfo {author} {\bibfnamefont {E.}~\bibnamefont {Bimbard}},
  \bibinfo {author} {\bibfnamefont {J.}~\bibnamefont {Stanojevic}}, \bibinfo
  {author} {\bibfnamefont {A.~J.}\ \bibnamefont {Hilliard}}, \bibinfo {author}
  {\bibfnamefont {F.}~\bibnamefont {Nogrette}}, \bibinfo {author}
  {\bibfnamefont {R.}~\bibnamefont {Tualle-Brouri}}, \bibinfo {author}
  {\bibfnamefont {A.}~\bibnamefont {Ourjoumtsev}}, \ and\ \bibinfo {author}
  {\bibfnamefont {P.}~\bibnamefont {Grangier}},\ }\href {\doibase
  10.1109/CLEOE-IQEC.2013.6801609} {\bibfield  {journal} {\bibinfo  {journal}
  {Conf. CLEO/Europe-IQEC 2013}\ }\textbf {\bibinfo {volume} {109}},\ \bibinfo
  {pages} {233602} (\bibinfo {year} {2013})}\BibitemShut {NoStop}%
\bibitem [{\citenamefont {Ningyuan}\ \emph {et~al.}(2016)\citenamefont
  {Ningyuan}, \citenamefont {Georgakopoulos}, \citenamefont {Ryou},
  \citenamefont {Schine}, \citenamefont {Sommer},\ and\ \citenamefont
  {Simon}}]{SimonPRA2015}%
  \BibitemOpen
  \bibfield  {author} {\bibinfo {author} {\bibfnamefont {J.}~\bibnamefont
  {Ningyuan}}, \bibinfo {author} {\bibfnamefont {A.}~\bibnamefont
  {Georgakopoulos}}, \bibinfo {author} {\bibfnamefont {A.}~\bibnamefont
  {Ryou}}, \bibinfo {author} {\bibfnamefont {N.}~\bibnamefont {Schine}},
  \bibinfo {author} {\bibfnamefont {A.}~\bibnamefont {Sommer}}, \ and\ \bibinfo
  {author} {\bibfnamefont {J.}~\bibnamefont {Simon}},\ }\href {\doibase
  10.1103/PhysRevA.93.041802} {\bibfield  {journal} {\bibinfo  {journal} {Phys.
  Rev. A}\ }\textbf {\bibinfo {volume} {93}},\ \bibinfo {pages} {041802}
  (\bibinfo {year} {2016})}\BibitemShut {NoStop}%
\bibitem [{\citenamefont {Roch}\ \emph {et~al.}(1997)\citenamefont {Roch},
  \citenamefont {Vigneron}, \citenamefont {Grelu}, \citenamefont {Sinatra},
  \citenamefont {Poizat},\ and\ \citenamefont {Grangier}}]{Grangier1997}%
  \BibitemOpen
  \bibfield  {author} {\bibinfo {author} {\bibfnamefont {J.-F.}\ \bibnamefont
  {Roch}}, \bibinfo {author} {\bibfnamefont {K.}~\bibnamefont {Vigneron}},
  \bibinfo {author} {\bibfnamefont {P.}~\bibnamefont {Grelu}}, \bibinfo
  {author} {\bibfnamefont {A.}~\bibnamefont {Sinatra}}, \bibinfo {author}
  {\bibfnamefont {J.-P.}\ \bibnamefont {Poizat}}, \ and\ \bibinfo {author}
  {\bibfnamefont {P.}~\bibnamefont {Grangier}},\ }\href {\doibase
  10.1103/PhysRevLett.78.634} {\bibfield  {journal} {\bibinfo  {journal} {Phys.
  Rev. Lett.}\ }\textbf {\bibinfo {volume} {78}},\ \bibinfo {pages} {634}
  (\bibinfo {year} {1997})}\BibitemShut {NoStop}%
\bibitem [{\citenamefont {Gorshkov}\ \emph {et~al.}(2007)\citenamefont
  {Gorshkov}, \citenamefont {Andr{\'{e}}}, \citenamefont {Lukin},\ and\
  \citenamefont {S{\o}rensen}}]{GorshkovPRA2007}%
  \BibitemOpen
  \bibfield  {author} {\bibinfo {author} {\bibfnamefont {A.~V.}\ \bibnamefont
  {Gorshkov}}, \bibinfo {author} {\bibfnamefont {A.}~\bibnamefont
  {Andr{\'{e}}}}, \bibinfo {author} {\bibfnamefont {M.~D.}\ \bibnamefont
  {Lukin}}, \ and\ \bibinfo {author} {\bibfnamefont {A.~S.}\ \bibnamefont
  {S{\o}rensen}},\ }\href {\doibase 10.1103/PhysRevA.76.033806} {\bibfield
  {journal} {\bibinfo  {journal} {Phys. Rev. A}\ }\textbf {\bibinfo {volume}
  {76}},\ \bibinfo {pages} {033806} (\bibinfo {year} {2007})}\BibitemShut
  {NoStop}%
\bibitem [{\citenamefont {Das}\ \emph {et~al.}(2016)\citenamefont {Das},
  \citenamefont {Grankin}, \citenamefont {Iakoupov}, \citenamefont {Brion},
  \citenamefont {Borregaard}, \citenamefont {Boddeda}, \citenamefont {Usmani},
  \citenamefont {Ourjoumtsev}, \citenamefont {Grangier},\ and\ \citenamefont
  {S{\o}rensen}}]{Sorenso2016RydCav}%
  \BibitemOpen
  \bibfield  {author} {\bibinfo {author} {\bibfnamefont {S.}~\bibnamefont
  {Das}}, \bibinfo {author} {\bibfnamefont {A.}~\bibnamefont {Grankin}},
  \bibinfo {author} {\bibfnamefont {I.}~\bibnamefont {Iakoupov}}, \bibinfo
  {author} {\bibfnamefont {E.}~\bibnamefont {Brion}}, \bibinfo {author}
  {\bibfnamefont {J.}~\bibnamefont {Borregaard}}, \bibinfo {author}
  {\bibfnamefont {R.}~\bibnamefont {Boddeda}}, \bibinfo {author} {\bibfnamefont
  {I.}~\bibnamefont {Usmani}}, \bibinfo {author} {\bibfnamefont
  {A.}~\bibnamefont {Ourjoumtsev}}, \bibinfo {author} {\bibfnamefont
  {P.}~\bibnamefont {Grangier}}, \ and\ \bibinfo {author} {\bibfnamefont
  {A.~S.}\ \bibnamefont {S{\o}rensen}},\ }\href {\doibase
  10.1103/PhysRevA.93.040303} {\bibfield  {journal} {\bibinfo  {journal} {Phys.
  Rev. A}\ }\textbf {\bibinfo {volume} {93}},\ \bibinfo {pages} {040303}
  (\bibinfo {year} {2016})}\BibitemShut {NoStop}%
\bibitem [{\citenamefont {Hao}\ \emph {et~al.}(2015)\citenamefont {Hao},
  \citenamefont {Lin}, \citenamefont {Xia}, \citenamefont {Lin}, \citenamefont
  {Niu},\ and\ \citenamefont {Gong}}]{Gong2015}%
  \BibitemOpen
  \bibfield  {author} {\bibinfo {author} {\bibfnamefont {Y.~M.}\ \bibnamefont
  {Hao}}, \bibinfo {author} {\bibfnamefont {G.~W.}\ \bibnamefont {Lin}},
  \bibinfo {author} {\bibfnamefont {K.}~\bibnamefont {Xia}}, \bibinfo {author}
  {\bibfnamefont {X.~M.}\ \bibnamefont {Lin}}, \bibinfo {author} {\bibfnamefont
  {Y.~P.}\ \bibnamefont {Niu}}, \ and\ \bibinfo {author} {\bibfnamefont
  {S.~Q.}\ \bibnamefont {Gong}},\ }\href {\doibase 10.1038/srep10005}
  {\bibfield  {journal} {\bibinfo  {journal} {Sci. Rep.}\ }\textbf {\bibinfo
  {volume} {5}},\ \bibinfo {pages} {10005} (\bibinfo {year}
  {2015})}\BibitemShut {NoStop}%
\bibitem [{\citenamefont {Iakoupov}\ \emph
  {et~al.}(2016{\natexlab{a}})\citenamefont {Iakoupov}, \citenamefont
  {Borregaard},\ and\ \citenamefont {S�rensen}}]{SorensonArxiv2016Sagnac}%
  \BibitemOpen
  \bibfield  {author} {\bibinfo {author} {\bibfnamefont {I.}~\bibnamefont
  {Iakoupov}}, \bibinfo {author} {\bibfnamefont {J.}~\bibnamefont
  {Borregaard}}, \ and\ \bibinfo {author} {\bibfnamefont {A.~S.}\ \bibnamefont
  {S�rensen}},\ }\href {http://arxiv.org/abs/1610.09206} {\  (\bibinfo {year}
  {2016}{\natexlab{a}})},\ \Eprint {http://arxiv.org/abs/1610.09206}
  {arXiv:1610.09206} \BibitemShut {NoStop}%
\bibitem [{\citenamefont {Gorshkov}\ \emph {et~al.}(2011)\citenamefont
  {Gorshkov}, \citenamefont {Otterbach}, \citenamefont {Fleischhauer},
  \citenamefont {Pohl},\ and\ \citenamefont {Lukin}}]{Gorshkov2011a}%
  \BibitemOpen
  \bibfield  {author} {\bibinfo {author} {\bibfnamefont {A.~V.}\ \bibnamefont
  {Gorshkov}}, \bibinfo {author} {\bibfnamefont {J.}~\bibnamefont {Otterbach}},
  \bibinfo {author} {\bibfnamefont {M.}~\bibnamefont {Fleischhauer}}, \bibinfo
  {author} {\bibfnamefont {T.}~\bibnamefont {Pohl}}, \ and\ \bibinfo {author}
  {\bibfnamefont {M.~D.}\ \bibnamefont {Lukin}},\ }\href {\doibase
  10.1103/PhysRevLett.107.133602} {\bibfield  {journal} {\bibinfo  {journal}
  {Phys. Rev. Lett.}\ }\textbf {\bibinfo {volume} {107}},\ \bibinfo {pages} {1}
  (\bibinfo {year} {2011})}\BibitemShut {NoStop}%
\bibitem [{\citenamefont {Fleischhauer}\ \emph {et~al.}(2005)\citenamefont
  {Fleischhauer}, \citenamefont {Imamoglu},\ and\ \citenamefont
  {Marangos}}]{Fleischhauer2005EITRev}%
  \BibitemOpen
  \bibfield  {author} {\bibinfo {author} {\bibfnamefont {M.}~\bibnamefont
  {Fleischhauer}}, \bibinfo {author} {\bibfnamefont {A.}~\bibnamefont
  {Imamoglu}}, \ and\ \bibinfo {author} {\bibfnamefont {J.~P.}\ \bibnamefont
  {Marangos}},\ }\href {\doibase 10.1103/RevModPhys.77.633} {\bibfield
  {journal} {\bibinfo  {journal} {Rev. Mod. Phys.}\ }\textbf {\bibinfo {volume}
  {77}},\ \bibinfo {pages} {633} (\bibinfo {year} {2005})}\BibitemShut
  {NoStop}%
\bibitem [{\citenamefont {Reinhard}\ \emph {et~al.}(2007)\citenamefont
  {Reinhard}, \citenamefont {Liebisch}, \citenamefont {Knuffman},\ and\
  \citenamefont {Raithel}}]{Raithel2007}%
  \BibitemOpen
  \bibfield  {author} {\bibinfo {author} {\bibfnamefont {A.}~\bibnamefont
  {Reinhard}}, \bibinfo {author} {\bibfnamefont {T.~C.}\ \bibnamefont
  {Liebisch}}, \bibinfo {author} {\bibfnamefont {B.}~\bibnamefont {Knuffman}},
  \ and\ \bibinfo {author} {\bibfnamefont {G.}~\bibnamefont {Raithel}},\ }\href
  {\doibase 10.1103/PhysRevA.75.032712} {\bibfield  {journal} {\bibinfo
  {journal} {Phys. Rev. A}\ }\textbf {\bibinfo {volume} {75}},\ \bibinfo
  {pages} {032712} (\bibinfo {year} {2007})}\BibitemShut {NoStop}%
\bibitem [{\citenamefont {Paspalakis}\ and\ \citenamefont
  {Knight}(2002)}]{Paspalakis2002}%
  \BibitemOpen
  \bibfield  {author} {\bibinfo {author} {\bibfnamefont {E.}~\bibnamefont
  {Paspalakis}}\ and\ \bibinfo {author} {\bibfnamefont {P.~L.}\ \bibnamefont
  {Knight}},\ }\href {\doibase 10.1103/PhysRevA.66.015802} {\bibfield
  {journal} {\bibinfo  {journal} {Phys. Rev. A}\ }\textbf {\bibinfo {volume}
  {66}},\ \bibinfo {pages} {158021} (\bibinfo {year} {2002})}\BibitemShut
  {NoStop}%
\bibitem [{Note1()}]{Note1}%
  \BibitemOpen
  \bibinfo {note} {\label {note1} In fact, all multi-level EIT susceptibilities
  $\chi =\chi _1[1-\DOTSB \sum@ \slimits@ _n \Omega _n^2/ (\delta _n+i\gamma
  _n)/(\Delta +i\Gamma )]^{-1}$ with multiple Rabi frequencies $\Omega _n$ and
  detunings $\delta _n$ \cite {Paspalakis2002} form the same circle when
  $\gamma _n=0$.}\BibitemShut {Stop}%
\bibitem [{\citenamefont {Waks}\ and\ \citenamefont
  {Vuckovic}(2006)}]{Waks2006}%
  \BibitemOpen
  \bibfield  {author} {\bibinfo {author} {\bibfnamefont {E.}~\bibnamefont
  {Waks}}\ and\ \bibinfo {author} {\bibfnamefont {J.}~\bibnamefont
  {Vuckovic}},\ }\href {\doibase 10.1103/PhysRevA.73.041803} {\bibfield
  {journal} {\bibinfo  {journal} {Phys. Rev. A}\ }\textbf {\bibinfo {volume}
  {73}},\ \bibinfo {pages} {13} (\bibinfo {year} {2006})}\BibitemShut {NoStop}%
\bibitem [{\citenamefont {Shomroni}\ \emph {et~al.}(2014)\citenamefont
  {Shomroni}, \citenamefont {Rosenblum}, \citenamefont {Lovsky}, \citenamefont
  {Bechler}, \citenamefont {Guendelman},\ and\ \citenamefont
  {Dayan}}]{Dayan2014}%
  \BibitemOpen
  \bibfield  {author} {\bibinfo {author} {\bibfnamefont {I.}~\bibnamefont
  {Shomroni}}, \bibinfo {author} {\bibfnamefont {S.}~\bibnamefont {Rosenblum}},
  \bibinfo {author} {\bibfnamefont {Y.}~\bibnamefont {Lovsky}}, \bibinfo
  {author} {\bibfnamefont {O.}~\bibnamefont {Bechler}}, \bibinfo {author}
  {\bibfnamefont {G.}~\bibnamefont {Guendelman}}, \ and\ \bibinfo {author}
  {\bibfnamefont {B.}~\bibnamefont {Dayan}},\ }\href {\doibase
  10.1126/science.1254699} {\bibfield  {journal} {\bibinfo  {journal}
  {Science}\ }\textbf {\bibinfo {volume} {345}},\ \bibinfo {pages} {903}
  (\bibinfo {year} {2014})}\BibitemShut {NoStop}%
\bibitem [{\citenamefont {Ritter}\ \emph {et~al.}(2016)\citenamefont {Ritter},
  \citenamefont {Gruhler}, \citenamefont {Pernice}, \citenamefont {K\"{u}bler},
  \citenamefont {Pfau},\ and\ \citenamefont {L\"{o}w}}]{Pfau2016Ring}%
  \BibitemOpen
  \bibfield  {author} {\bibinfo {author} {\bibfnamefont {R.}~\bibnamefont
  {Ritter}}, \bibinfo {author} {\bibfnamefont {N.}~\bibnamefont {Gruhler}},
  \bibinfo {author} {\bibfnamefont {W.~H.~P.}\ \bibnamefont {Pernice}},
  \bibinfo {author} {\bibfnamefont {H.}~\bibnamefont {K\"{u}bler}}, \bibinfo
  {author} {\bibfnamefont {T.}~\bibnamefont {Pfau}}, \ and\ \bibinfo {author}
  {\bibfnamefont {R.}~\bibnamefont {L\"{o}w}},\ }\href {\doibase
  10.1088/1367-2630/18/10/103031} {\bibfield  {journal} {\bibinfo  {journal}
  {New J. Phys.}\ }\textbf {\bibinfo {volume} {18}},\ \bibinfo {pages} {103031}
  (\bibinfo {year} {2016})}\BibitemShut {NoStop}%
\bibitem [{\citenamefont {{A. Yariv and P. Yeh}}(2006)}]{Yariv2007}%
  \BibitemOpen
  \bibfield  {author} {\bibinfo {author} {\bibnamefont {{A. Yariv and P.
  Yeh}}},\ }\href@noop {} {\emph {\bibinfo {title} {{Photonics: Optical
  Electronics in Modern Communications}}}}\ (\bibinfo  {publisher} {New York:
  oxford university press},\ \bibinfo {year} {2006})\BibitemShut {NoStop}%
\bibitem [{\citenamefont {Siegman}(1986)}]{Siegman1986}%
  \BibitemOpen
  \bibfield  {author} {\bibinfo {author} {\bibfnamefont {A.~E.}\ \bibnamefont
  {Siegman}},\ }\href@noop {} {\emph {\bibinfo {title} {{Lasers}}}}\ (\bibinfo
  {publisher} {University Science Books},\ \bibinfo {year} {1986})\BibitemShut
  {NoStop}%
\bibitem [{Note2()}]{Note2}%
  \BibitemOpen
  \bibinfo {note} {For example, $|\protect \qopname \relax o{arg}(r_0)-\protect
  \qopname \relax o{arg}(r_1)|< \pi /2$ and $|\protect \qopname \relax
  o{arg}(t_0)-\protect \qopname \relax o{arg}(t_1)|< \pi /2$ within the
  resonance linewidth $|\protect \mathrm {Re}(\theta _0-\theta _1)|<\pi
  /\protect \mathcal {F}$.}\BibitemShut {Stop}%
\bibitem [{\citenamefont {Erdogan}(1997)}]{Erdogan1997}%
  \BibitemOpen
  \bibfield  {author} {\bibinfo {author} {\bibfnamefont {T.}~\bibnamefont
  {Erdogan}},\ }\href {\doibase 10.1109/50.618322} {\bibfield  {journal}
  {\bibinfo  {journal} {J. Light. Technol.}\ }\textbf {\bibinfo {volume}
  {15}},\ \bibinfo {pages} {1277} (\bibinfo {year} {1997})}\BibitemShut
  {NoStop}%
\bibitem [{\citenamefont {Zimmer}\ \emph {et~al.}(2006)\citenamefont {Zimmer},
  \citenamefont {Andr{\'{e}}}, \citenamefont {Lukin},\ and\ \citenamefont
  {Fleischhauer}}]{Zimmer2006}%
  \BibitemOpen
  \bibfield  {author} {\bibinfo {author} {\bibfnamefont {F.~E.}\ \bibnamefont
  {Zimmer}}, \bibinfo {author} {\bibfnamefont {A.}~\bibnamefont {Andr{\'{e}}}},
  \bibinfo {author} {\bibfnamefont {M.~D.}\ \bibnamefont {Lukin}}, \ and\
  \bibinfo {author} {\bibfnamefont {M.}~\bibnamefont {Fleischhauer}},\ }\href
  {\doibase 10.1016/j.optcom.2006.03.075} {\bibfield  {journal} {\bibinfo
  {journal} {Opt. Commun.}\ }\textbf {\bibinfo {volume} {264}},\ \bibinfo
  {pages} {441} (\bibinfo {year} {2006})}\BibitemShut {NoStop}%
\bibitem [{SM()}]{SM}%
  \BibitemOpen
  \href@noop {} {\bibinfo  {journal} {See Supplemental Material at :URL: for
  more details on the parameter optimization}\ }\BibitemShut {NoStop}%
\bibitem [{\citenamefont {Bajcsy}\ \emph {et~al.}(2003)\citenamefont {Bajcsy},
  \citenamefont {Zibrov},\ and\ \citenamefont {Lukin}}]{Bajcsy2003a}%
  \BibitemOpen
\bibfield  {journal} {  }\bibfield  {author} {\bibinfo {author} {\bibfnamefont
  {M.}~\bibnamefont {Bajcsy}}, \bibinfo {author} {\bibfnamefont {A.~S.}\
  \bibnamefont {Zibrov}}, \ and\ \bibinfo {author} {\bibfnamefont {M.~D.}\
  \bibnamefont {Lukin}},\ }\href {\doibase 10.1038/nature02176} {\bibfield
  {journal} {\bibinfo  {journal} {Nature}\ }\textbf {\bibinfo {volume} {426}},\
  \bibinfo {pages} {638} (\bibinfo {year} {2003})}\BibitemShut {NoStop}%
\bibitem [{\citenamefont {Iakoupov}\ \emph
  {et~al.}(2016{\natexlab{b}})\citenamefont {Iakoupov}, \citenamefont {Ott},
  \citenamefont {Chang},\ and\ \citenamefont {S{\o}rensen}}]{Sorenson2017}%
  \BibitemOpen
  \bibfield  {author} {\bibinfo {author} {\bibfnamefont {I.}~\bibnamefont
  {Iakoupov}}, \bibinfo {author} {\bibfnamefont {J.~R.}\ \bibnamefont {Ott}},
  \bibinfo {author} {\bibfnamefont {D.~E.}\ \bibnamefont {Chang}}, \ and\
  \bibinfo {author} {\bibfnamefont {A.~S.}\ \bibnamefont {S{\o}rensen}},\
  }\href {\doibase 10.1103/PhysRevA.94.053824} {\bibfield  {journal} {\bibinfo
  {journal} {Phys. Rev. A}\ }\textbf {\bibinfo {volume} {94}},\ \bibinfo
  {pages} {053824} (\bibinfo {year} {2016}{\natexlab{b}})}\BibitemShut
  {NoStop}%
\bibitem [{\citenamefont {Boyd}(2012)}]{Boyd2011}%
  \BibitemOpen
  \bibfield  {author} {\bibinfo {author} {\bibfnamefont {R.~W.}\ \bibnamefont
  {Boyd}},\ }\href {\doibase 10.1364/JOSAB.29.002644} {\bibfield  {journal}
  {\bibinfo  {journal} {J. Opt. Soc. Am. B}\ }\textbf {\bibinfo {volume}
  {29}},\ \bibinfo {pages} {2644} (\bibinfo {year} {2012})}\BibitemShut
  {NoStop}%
\bibitem [{\citenamefont {Little}\ \emph {et~al.}(2013)\citenamefont {Little},
  \citenamefont {Starling}, \citenamefont {Howell}, \citenamefont {Cohen},
  \citenamefont {Shwa},\ and\ \citenamefont {Katz}}]{KatzHowell2013}%
  \BibitemOpen
  \bibfield  {author} {\bibinfo {author} {\bibfnamefont {B.}~\bibnamefont
  {Little}}, \bibinfo {author} {\bibfnamefont {D.~J.}\ \bibnamefont
  {Starling}}, \bibinfo {author} {\bibfnamefont {J.~C.}\ \bibnamefont
  {Howell}}, \bibinfo {author} {\bibfnamefont {R.~D.}\ \bibnamefont {Cohen}},
  \bibinfo {author} {\bibfnamefont {D.}~\bibnamefont {Shwa}}, \ and\ \bibinfo
  {author} {\bibfnamefont {N.}~\bibnamefont {Katz}},\ }\href {\doibase
  10.1103/PhysRevA.87.043815} {\bibfield  {journal} {\bibinfo  {journal} {Phys.
  Rev. A}\ }\textbf {\bibinfo {volume} {87}},\ \bibinfo {pages} {2} (\bibinfo
  {year} {2013})}\BibitemShut {NoStop}%
\bibitem [{\citenamefont {Lin}\ \emph {et~al.}(2009)\citenamefont {Lin},
  \citenamefont {Liao}, \citenamefont {Peters}, \citenamefont {Chou},
  \citenamefont {Wang}, \citenamefont {Cho}, \citenamefont {Kuan},\ and\
  \citenamefont {Yu}}]{Lin2009}%
  \BibitemOpen
  \bibfield  {author} {\bibinfo {author} {\bibfnamefont {Y.~W.}\ \bibnamefont
  {Lin}}, \bibinfo {author} {\bibfnamefont {W.~T.}\ \bibnamefont {Liao}},
  \bibinfo {author} {\bibfnamefont {T.}~\bibnamefont {Peters}}, \bibinfo
  {author} {\bibfnamefont {H.~C.}\ \bibnamefont {Chou}}, \bibinfo {author}
  {\bibfnamefont {J.~S.}\ \bibnamefont {Wang}}, \bibinfo {author}
  {\bibfnamefont {H.~W.}\ \bibnamefont {Cho}}, \bibinfo {author} {\bibfnamefont
  {P.~C.}\ \bibnamefont {Kuan}}, \ and\ \bibinfo {author} {\bibfnamefont
  {I.~A.}\ \bibnamefont {Yu}},\ }\href {\doibase
  10.1103/PhysRevLett.102.213601} {\bibfield  {journal} {\bibinfo  {journal}
  {Phys. Rev. Lett.}\ }\textbf {\bibinfo {volume} {102}},\ \bibinfo {pages}
  {213601} (\bibinfo {year} {2009})}\BibitemShut {NoStop}%
\bibitem [{\citenamefont {Sparkes}\ \emph {et~al.}(2013)\citenamefont
  {Sparkes}, \citenamefont {Bernu}, \citenamefont {Hosseini}, \citenamefont
  {Geng}, \citenamefont {Glorieux}, \citenamefont {Altin}, \citenamefont {Lam},
  \citenamefont {Robins},\ and\ \citenamefont {Buchler}}]{Sparkes2013a}%
  \BibitemOpen
  \bibfield  {author} {\bibinfo {author} {\bibfnamefont {B.~M.}\ \bibnamefont
  {Sparkes}}, \bibinfo {author} {\bibfnamefont {J.}~\bibnamefont {Bernu}},
  \bibinfo {author} {\bibfnamefont {M.}~\bibnamefont {Hosseini}}, \bibinfo
  {author} {\bibfnamefont {J.}~\bibnamefont {Geng}}, \bibinfo {author}
  {\bibfnamefont {Q.}~\bibnamefont {Glorieux}}, \bibinfo {author}
  {\bibfnamefont {P.~A.}\ \bibnamefont {Altin}}, \bibinfo {author}
  {\bibfnamefont {P.~K.}\ \bibnamefont {Lam}}, \bibinfo {author} {\bibfnamefont
  {N.~P.}\ \bibnamefont {Robins}}, \ and\ \bibinfo {author} {\bibfnamefont
  {B.~C.}\ \bibnamefont {Buchler}},\ }\href {\doibase
  10.1088/1367-2630/15/8/085027} {\bibfield  {journal} {\bibinfo  {journal}
  {New J. Phys.}\ }\textbf {\bibinfo {volume} {15}},\ \bibinfo {pages} {085027}
  (\bibinfo {year} {2013})}\BibitemShut {NoStop}%
\bibitem [{\citenamefont {Blatt}\ \emph {et~al.}(2014)\citenamefont {Blatt},
  \citenamefont {Halfmann},\ and\ \citenamefont {Peters}}]{Blatt2014}%
  \BibitemOpen
  \bibfield  {author} {\bibinfo {author} {\bibfnamefont {F.}~\bibnamefont
  {Blatt}}, \bibinfo {author} {\bibfnamefont {T.}~\bibnamefont {Halfmann}}, \
  and\ \bibinfo {author} {\bibfnamefont {T.}~\bibnamefont {Peters}},\ }\href
  {\doibase 10.1364/OL.39.000446} {\bibfield  {journal} {\bibinfo  {journal}
  {Optics Letters}\ }\textbf {\bibinfo {volume} {39}},\ \bibinfo {pages} {446}
  (\bibinfo {year} {2014})}\BibitemShut {NoStop}%
\bibitem [{\citenamefont {Kaczmarek}\ \emph {et~al.}(2015)\citenamefont
  {Kaczmarek}, \citenamefont {Saunders}, \citenamefont {Sprague}, \citenamefont
  {Kolthammer}, \citenamefont {Feizpour}, \citenamefont {Ledingham},
  \citenamefont {Brecht}, \citenamefont {Poem}, \citenamefont {Walmsley},\ and\
  \citenamefont {Nunn}}]{Walmsley2015OD}%
  \BibitemOpen
  \bibfield  {author} {\bibinfo {author} {\bibfnamefont {K.~T.}\ \bibnamefont
  {Kaczmarek}}, \bibinfo {author} {\bibfnamefont {D.~J.}\ \bibnamefont
  {Saunders}}, \bibinfo {author} {\bibfnamefont {M.~R.}\ \bibnamefont
  {Sprague}}, \bibinfo {author} {\bibfnamefont {W.~S.}\ \bibnamefont
  {Kolthammer}}, \bibinfo {author} {\bibfnamefont {A.}~\bibnamefont
  {Feizpour}}, \bibinfo {author} {\bibfnamefont {P.~M.}\ \bibnamefont
  {Ledingham}}, \bibinfo {author} {\bibfnamefont {B.}~\bibnamefont {Brecht}},
  \bibinfo {author} {\bibfnamefont {E.}~\bibnamefont {Poem}}, \bibinfo {author}
  {\bibfnamefont {I.~A.}\ \bibnamefont {Walmsley}}, \ and\ \bibinfo {author}
  {\bibfnamefont {J.}~\bibnamefont {Nunn}},\ }\href {\doibase
  10.1364/OL.40.005582} {\bibfield  {journal} {\bibinfo  {journal} {Opt.
  Lett.}\ }\textbf {\bibinfo {volume} {40}},\ \bibinfo {pages} {005582}
  (\bibinfo {year} {2015})}\BibitemShut {NoStop}%
\end{thebibliography}

%

\end{document}


\title{Induced cavities for photonic quantum gates: \\ Supplemental Material}
\author{Ohr Lahad}
\affiliation{Department of Physics of Complex Systems, Weizmann Institute of Science, Rehovot 76100, Israel}
\author{Ofer Firstenberg}
\affiliation{Department of Physics of Complex Systems, Weizmann Institute of Science, Rehovot 76100, Israel}

\maketitle

\beginsupplement

\section{Induced Bragg grating}

This section elaborates on the performance and the parameters of a Bragg cavity induced in an atomic medium by spatial modulation of a two-photon resonance.

\subsection{Relations between the Bragg coefficients $\sigma L$, $\kappa L$ and the atomic parameters $x$, $y$, $\Delta/\Gamma$ }
\begin{figure} 
\centering
\includegraphics[scale=0.6,trim=0cm 0cm 0cm 0cm,clip=true]{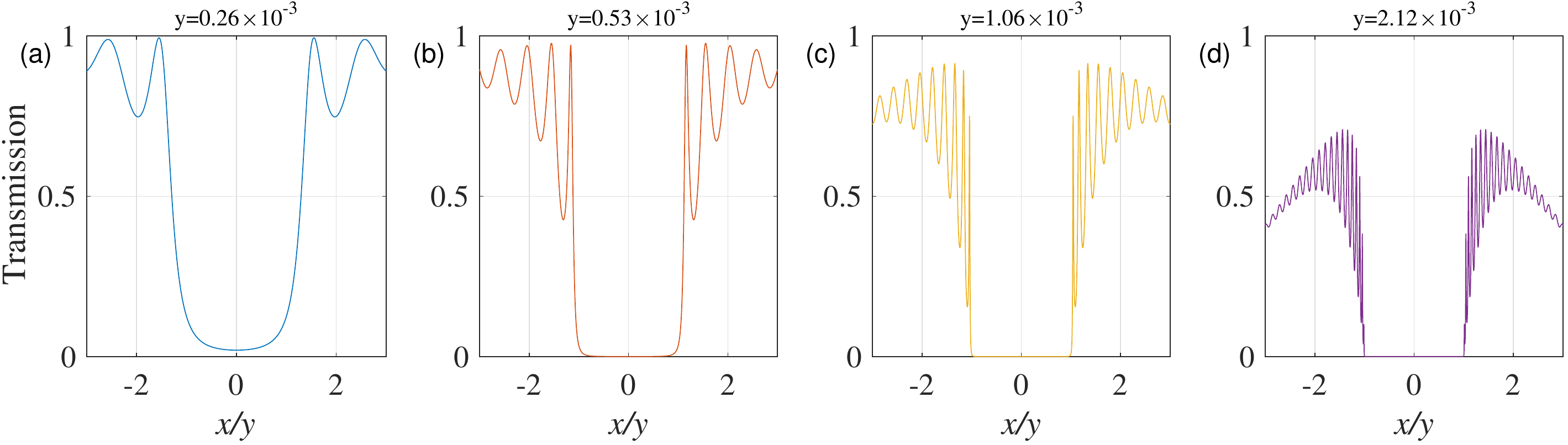}
\caption{Transmission spectra for $d=10^4$ and different $y$ values. The bandedge is located at $x\approx y$. 
As $y$ is increased, the transmission peaks become narrower (and thus the intensity build-up inside the grating is enhanced). However, increasing $y$ also degrades the transmission at the resonances due to the quadratic dependence of absorption on the two-photon detuning.
}\label{fig2}
\end{figure}

\begin{figure} 
\centering
\includegraphics[scale=0.5,trim=0cm 0cm 0cm 0cm,clip=true]{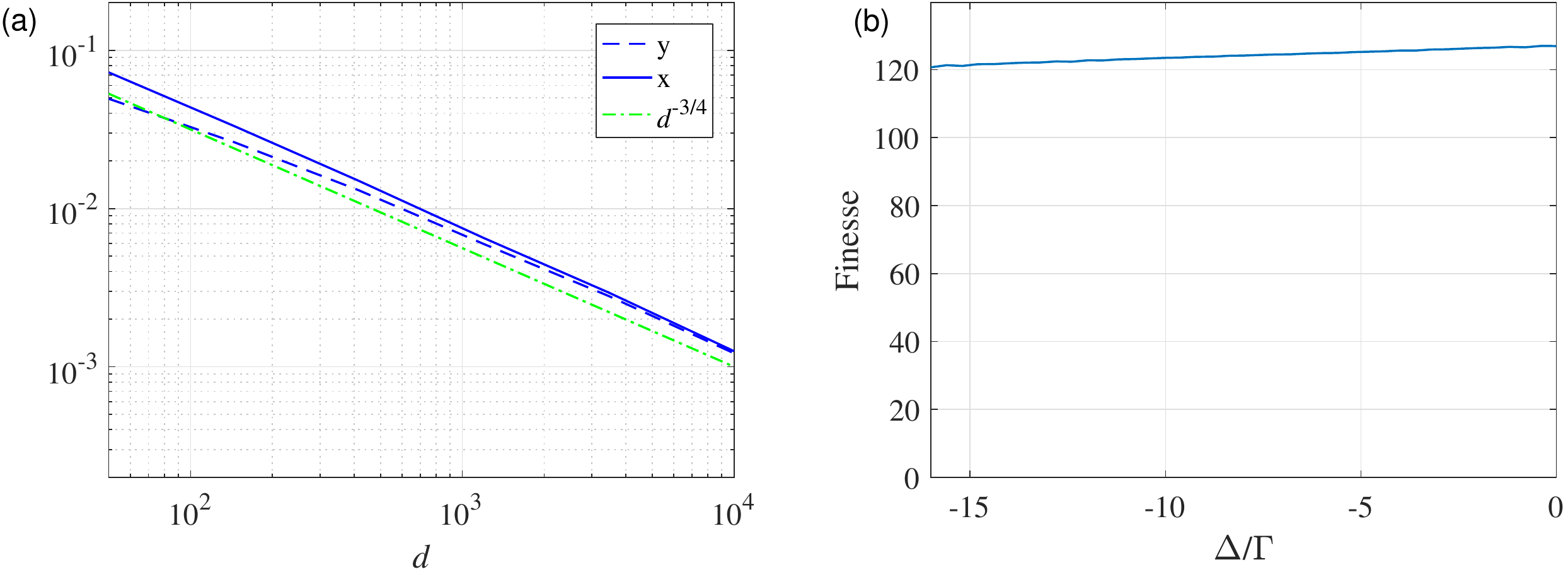}
\caption{Numerical optimization of the parameters $x$ and $y$ for maximizing the finesse of the induced cavity. (a) The two parameters scale as $d^{-3/4}$ [green dash-dot, see Eq.~(\ref{eqn:yd})] and approach each other for large $d$. (b) Weak dependence of the finesse on $\Delta$ for $d=10^4$. 
}\label{fig1}
\end{figure}

The scheme we analyze relies on the longitudinal modulation of the refractive index, owing to the steep linear dispersion around the two-photon resonance under EIT conditions. We characterize the modulation using the parameters $x=\delta\Gamma/\Omega^2$ and $y=\Delta_s\Gamma/(2\Omega^2)$. The resulting Bragg coefficients $\sigma L$ and $\kappa L$ are given in Eq.~(10) of the main text, assuming the two-photon detuning is well within the EIT linewidth, $\delta_{\mathrm{mod}}\ll \Omega^2/ |\Delta+i\Gamma|$.  Both $x$ and $y$ contribute to the two-photon detuning $\delta_{\mathrm{mod}}$:
\begin{itemize}
\item A spatially-uniform detuning $\delta$ sets the mismatch $\sigma L \approx d x$ from the Bragg resonance.
\item A spatially-periodic light shift $\Delta_s \cos \left( 2 k_s z \right)$ determines the modulation depth $\kappa L \approx d y$ and sets the width of the bandgap and the finesse of the effective cavity.
\end{itemize}
It follows from $\delta_{\mathrm{mod}}\ll \Omega^2/ |\Delta+i\Gamma|$ that $x,y \ll 1$. As shown in Fig. \ref{fig2}, the transmission resonances first sharpen as $y$ is increased, but eventually degrade. The degradation at larger $y$ occurs due the finite EIT linewidth, introducing absorption proportional to $\delta_{\mathrm{mod}}^2$. As we will show in Sec.~\ref{sec2} [Eq.~(\ref{eqn:yd})], this sets in roughly at $y=d^{-3/4}$ for large $d$. Figure \ref{fig1}(a) shows the values of $x$ and $y$ optimized for obtaining the maximal finesse of the induced grating.

\hfill

The induced Bragg grating relies on the large group index --- the slope of the linear dispersion at EIT resonance --- for translating small frequency modulation to large refractive index modulation. Our analysis assumes that the EIT resonance is governed by power broadening, \emph{i.e.} that the metastable state (in our case, the Rydberg state) decays much slower than the EIT linewidth. In this regime, EIT has the peculiar property that \emph{the group index is independent of the one-photon detuning} $\Delta$, and it depends only on $\Omega^2/\Gamma$ and on the atomic density. For this reason, the grating performance is independent of $\Delta$ to leading order in the two-photon detunings $x$ and $y$. Therefore, as long as $\Delta/\Gamma \ll d^{3/4}$, it is safe to neglect its contribution in Eq.~(10) of the main text. We numerically verify this approximation, as shown in Fig.~\ref{fig1}(b). Note that the one-photon detuning $\Delta$ has an important influence on the performance of the phase gate, as it determines the susceptibility in the blockade volume; we discuss this in Sec.~\ref{sec2}.

\hfill

The first transmission resonance of the Bragg grating appears at $\lambda L=\sqrt{(\sigma L ) ^2 - (\kappa L ) ^2} = \pi \left( 1+ i \alpha \right) $. Substituting Eq.~(10) of the main text and keeping only leading orders in $x$ and $y$, we find
\begin{equation}\label{eqn:res_cond}
\frac{\pi}{d} \left(1 + i \alpha \right)=\sqrt{x^2-y^2 + 2ix^3}\approx \sqrt{x^2-y^2} +i \frac{ x^3} {\sqrt{x^2-y^2}}.
\end{equation}
Note that we further took $\Delta k=0$ to simplify the analysis; we verified numerically that its contribution is insignificant, as the effect of $\Delta k\ne0$ is largely equivalent to a small modification in $x$. The real part of Eq.~(\ref{eqn:res_cond}) sets the resonance condition, with the value of $x$ approaching that of $y$,
\begin{equation} \label{eqn:x_vs_y}
x=\sqrt{y^2+\left( \pi / d \right)^2},
\end{equation}
as shown in  Fig.~\ref{fig1}(a). The imaginary part of Eq.~(\ref{eqn:res_cond}), accounting for loss in the induced grating, is given by
\begin{equation} \label{eqn:alpha}
\alpha=\frac{d^2 x^3} {\pi^2}.
\end{equation}

\subsection{Finesse and resonant transmission of the induced grating}
As shown above by Eq.~(\ref{eqn:x_vs_y}) and Fig.~\ref{fig1}(a), at high optical depth, the ratio between the parameters $x$ and $y$ approaches unity $q\approx x/y \approx 1$. It follows from Eq.~(\ref{eqn:alpha}) that
\begin{equation}\label{eqn:alphay}
\alpha\approx  \frac{d^2 y^3} {\pi^2}
\end{equation}
and so
\begin{equation} \label{eqn:alphaz}
(d y)^2 = \sqrt{(d y)^4}=\sqrt{\alpha d y\frac{ (d y)^3}{\alpha}}=\pi \sqrt{\alpha d y}\sqrt{d}.
\end{equation}
Writing Eq.~(8) from the main text to leading order in $x$ and $y$, by substituting $\kappa L = dy$ and $q=1$, we get
\begin{equation} \label{eqn:sagTnB}
T\approx \left| \frac{1- \alpha \kappa L }{1+\alpha\kappa L} \right|^{2} \approx \left( \frac{1- \alpha d y }{1+\alpha d y} \right)^{2}
\textrm{~~~~and~~~~}
\FF  \approx   \frac{1}{\pi}  \left|\frac{2\kappa L}{1+\alpha\kappa L} \right|^2 \approx \frac{1}{\pi}  \left(\frac{2d y}{1+\alpha d y} \right)^2 \approx
\sqrt{d}  \frac{4\sqrt{\alpha d y}}{(1+\alpha d y)^2},
\end{equation}
where in the last step we used Eq.~(\ref{eqn:alphaz}). Identifying $w\equiv\alpha d y$ and using the identity
\begin{equation}
\underbrace{\frac{4\sqrt{w}}{(1+w)^2}}_{\FF/d}=\underbrace{\left( 1+ \frac{1-w}{1+w}\right)}_{1+\sqrt{T}} \underbrace{\sqrt{1-\left(\frac{1-w}{1+w}\right)^2}}_{\sqrt{1-T}},
\end{equation}
we arrive at the result given in the main text,
\begin{equation}\label{eqn:FandT}
\FF=(1+\sqrt{T})\sqrt{(1-T)d}.
\end{equation}

\section{Optimization of gate performance}\label{sec2}

The maximal finesse according to Eq.~(\ref{eqn:FandT}) is $\FF\approx 1.3 \sqrt{d}$, obtained for a transmission of $T=1/4$. For a phase gate, such low transmission is inapplicable, and indeed we find by full numerical optimization minimizing the gate loss, that for high optical depth, $T\approx 1$. For a high-fidelity gate, the scaling of $\FF$ with $d$ can be calculated by minimizing the gate loss
\begin{equation}\label{eqn:totaloss}
\frac{\FF^2}{4 d}+\frac{4\pi}{\FF\db}.
\end{equation}
This expression estimates the loss as the sum of the loss without the grating $\epsilon=(4\pi)/(\FF\db)$ [Eq.~(3) in the main text with $\phi=\pi$] and the loss due only to the grating $1-T\approx\FF^2/(4 d)$ [obtained from Eq.~(\ref{eqn:FandT}) by substituting $1+\sqrt{T}\approx 2$ assuming $1-T \ll 1$ ]. Minimizing (\ref{eqn:totaloss}) with respect to $\FF$ yields $\FF\approx\sqrt[3]{8\pi d /\db}$, which scales as $\sqrt[3]{d}$ rather than $\sqrt{d}$.

\hfill

\begin{figure}[b]
\centering
\includegraphics[scale=0.8,trim=0cm 0cm 0cm 0.cm,clip=true]{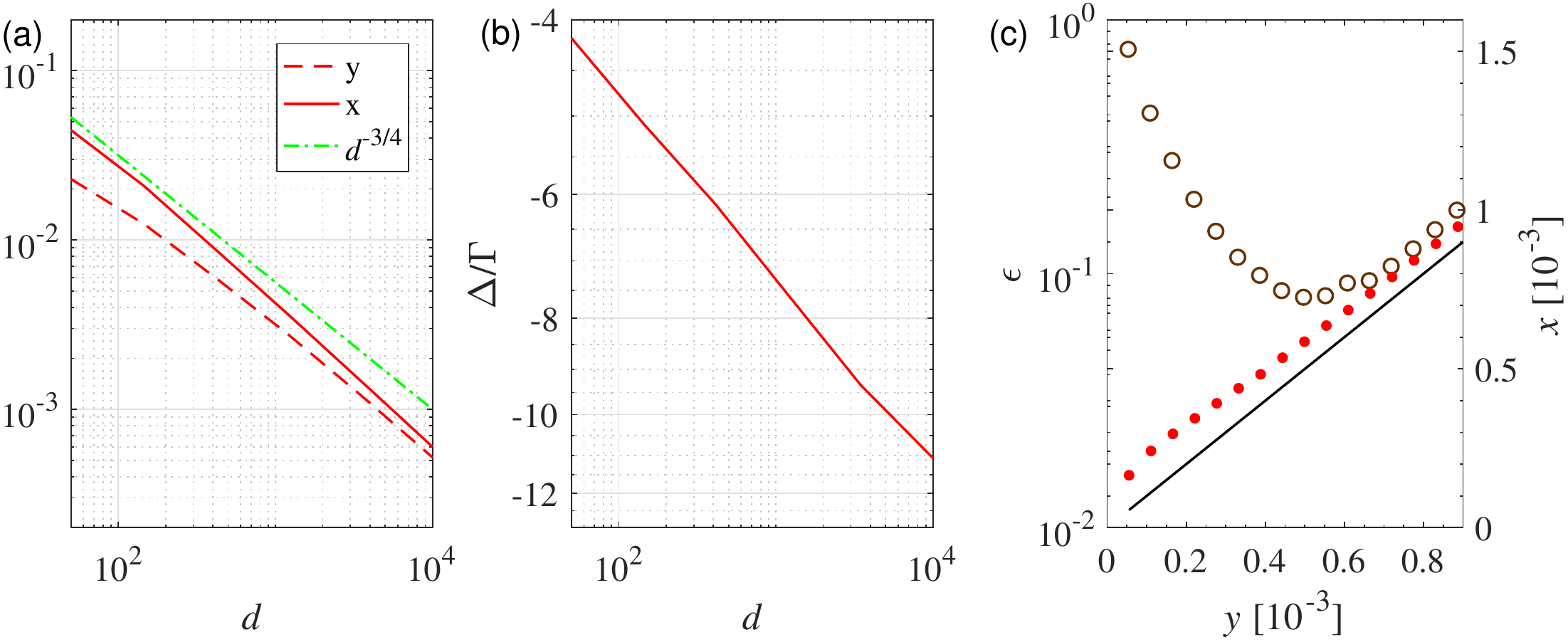}
\caption{Numerical optimization of the parameters $x$ and $y$ (a) and $\Delta/\Gamma$ (b), minimizing the loss of a conditional $\pi$ phase gate with $2\db=4\pi$. The green dash-dot line in (a) is $d^{-3/4}$, see Eq.~(\ref{eqn:yd}). (c) Dependence on $y$ for $d=10^4$. Brown circles (left axis): loss $\epsilon$  for a conditional $\pi$ phase gate. The minimal loss is obtained at $y\lesssim d^{-3/4}$, where the tradeoff between build-up due to the grating and atomic loss is optimal. Red dots (right axis): optimal values of $x$ when minimizing the loss $\epsilon$ for different $y$ values. The solid black line is $x=y$, shown for reference.
}\label{fig3}
\end{figure}

We can further study the regime $1-T \ll 1$ by going back to $T$ in Eq.~(\ref{eqn:sagTnB}) and solve for $\alpha d y$ up to first order in $(1-T)$:
\begin{equation} \label{eqn:ady}
\alpha d y=2\frac{1-\sqrt{1-(1-T)}}{1-T}-1\approx  \frac{1-T+(1-T)^2/4}{1-T}-1=\frac{1-T}{4}.
\end{equation}
Substituting Eq.~(\ref{eqn:alphay}) in the left hand side of Eq.~(\ref{eqn:ady}), we find $d^3 y^4 / \pi^2 =(1-T)/4$, or
\begin{equation} \label{eqn:yd}
y=\sqrt{\frac{\pi}{2}} (1-T)^{1/4} d^{-3/4}.
\end{equation}
The quartic root makes this relation largely insensitive to $T$, with the prefactor $\sqrt{\pi/2} (1-T)^{1/4}\approx 0.5-1$ for relevant values of $T$. As shown in Fig.~\ref{fig3}(a), the scaling of $x$ and $y$ found from exact numerical minimization of the gate loss agree with that estimated in Eq.~(\ref{eqn:yd}).


As already explained, the one-photon detuning $\Delta$ affects the phase gate performance by determining the atomic susceptibility (of the two-level system) in the blockade volume. The values of $\Delta/\Gamma$ obtained in the numerical optimization are shown in Fig.~\ref{fig3}(b). Some more insight into the optimization is given in Fig.~\ref{fig3}(c).

